\documentclass{pasj00}

\SetRunningHead{Y. Takeda et al.}{Stellar Parameters and 
Elemental Abundances of Late-G Giants}
\Received{2008/02/26}
\Accepted{2008/05/07}

\begin{document}

\title{Stellar Parameters and Elemental Abundances of Late-G Giants
\thanks{Based on observations carried out at Okayama Astrophysical 
Observatory (Okayama, Japan).}
\thanks{While the large datasets separately provided in the machine-readable 
form (electronic tables E1, E2, and E3) will be available in the electronic
edition of PASJ upon publication, they are also downloadable from the web site 
of $\langle$http://optik2.mtk.nao.ac.jp/\~{ }takeda/gg300/$\rangle$. }
}

\author{Yoichi \textsc{Takeda}}
\affil{National Astronomical Observatory, 2-21-1 Osawa, Mitaka, Tokyo 181-8588}
\email{takeda.yoichi@nao.ac.jp}

\author{Bun'ei \textsc{Sato}}
\affil{Tokyo Institute of Technology, 2-12-1 Ookayama, Meguro-ku, Tokyo 152-8550}
\email{sato.b.aa@m.titech.ac.jp}

\and

\author{Daisuke \textsc {Murata}}
\affil{Graduate School of Science and Technology, Kobe University, 
1-1 Rokkodai, Nada, Kobe 657-8501}
\email{dmurata@harbor.scitec.kobe-u.ac.jp}


\maketitle

\begin{abstract}
The properties of 322 intermediate-mass late-G giants (comprising 10 
planet-host stars) selected as the targets of Okayama Planet Search Program,
many of which are red-clump giants, were comprehensively investigated 
by establishing their various stellar parameters 
(atmospheric parameters including turbulent velocity fields, 
metallicity, luminosity, mass, age, projected rotational velocity, etc.),
and their photospheric chemical abundances for 17 elements,
in order to study their mutual dependence, connection with the existence 
of planets, and possible evolution-related characteristics. 
The metallicity distribution of planet-host giants was found to be 
almost the same as that of non-planet-host giants, making marked
contrast to the case of planet-host dwarfs tending to be metal-rich. 
Generally, the metallicities of these comparatively young (typical age of 
$\sim 10^{9}$~yr) giants tend to be somewhat lower than those of dwarfs at 
the same age, and super-metal-rich ([Fe/H] $>$ 0.2) giants appear to be lacking.
Apparent correlations were found between the abundances of C, O, and Na, 
suggesting that the surface compositions of these elements have undergone 
appreciable changes due to dredge-up of H-burning products by 
evolution-induced deep envelope mixing which becomes more efficient 
for higher-mass stars.
\end{abstract}

%

\newpage

\section{Introduction}

Since the beginning of the 21 century, a project of searching planets around
intermediate-mass (1.5--5 $M_{\odot}$) stars by using the Doppler technique 
has been undertaken at Okayama Astrophysical Observatory, 
which intensively targets evolved late-G type giants because they are 
considered to be most suitable for this purpose.\footnote{Intermediate-mass 
stars of other spectral types are less advantageous: spectral lines of B--F 
main-sequence stars are too few and broad/shallow to attain sufficient 
radial-velocity precision, while the atmospheres of cooler K giants tend 
to be intrinsically more unstable (compared to G-giants) which makes them 
less suitable for detecting delicate wobbles caused by orbiting planets.} 
This ``Okayama Planet Search Program'' has so far produced successful results
of newly discovering planets around 5 giants (HD~104985, $\epsilon$ Tau, 
18~Del, $\xi$~Aql, and HD~81688; cf. Sato et al. 2003, 2007, 2008) and 
1 brown dwarf around 11~Com (Liu et al. 2008). And it is still going on
with an extended monitoring sample of more than 300 stars (considerably 
increased from the first 50--60 targets when the project started), sorting 
out further new promising candidates of possible substellar companions,
which will be reported in forthcoming papers.

Now that such planet-host candidates have increasingly emerged from this 
project, it appears necessary to thoroughly review the characteristics of 
the sample targets, since such basic information consistently obtained 
for the whole sample is requisite to gain insight to the 
physical nature of planet-formation in intermediate-mass stars, given 
a number of questions to answer; e.g.: 
To which population do the sample stars belong in the Galaxy?
What are the key stellar parameters especially important to understand 
the mechanism of planet-formation (such as the mass, age, metallicity, 
rotational velocity, etc.)? Are there any difference between planet-host
giants and other normal giants? What are the ages of planet-host giants like?

Thus, as a natural extension of Takeda et al.'s (2005c; hereinafter referred 
to as Paper~I) study which confined to 57 late-G giants (the targets of 
the initial phase), we decided to conduct an extensive investigation for 
a total of 322 program stars, in order to clarify their properties from 
comprehensive point of view, so as to allow statistically meaningful 
discussion. Practically, our aim is to establish the atmospheric parameters, 
stellar fundamental quantities, kinematic parameters, and surface chemical 
compositions, while mostly based on the high-dispersion spectra accumulated 
during the course of the project. This is the primary purpose of this paper.

Besides, making use of the results we gained as by-products, we pay 
attention especially to discussing the chemical properties of these late-G 
giants, since several disputable tendencies were tentatively reported in 
Paper~I concerning the metallicity and the surface chemical composition 
(e.g., mass-dependent metallicity, subsolar trend of metallicity distribution, 
$\lambda$ Boo-like C vs. Si anti-correlation, under/over-abundance of O/Na
implying H-burning product dredged-up by non-canonical deep mixing).
Since the number of the sample has been considerably increased by a factor
of $\sim 6$, we would hope that more convincing results may be obtained
concerning the reality of these features, which is counted as another 
purpose of this study.

The remaining parts of this paper are organized as follows:
Section 2 describes the basic observational data invoked in this study.
The determinations of stellar parameters (atmospheric parameters by using 
the spectroscopic method, and fundamental parameters with the help of
theoretical stellar evolutionary tracks) are presented in section 3,
where comparisons with previous studies are also made. Section 4
deals with the kinematic parameters describing the orbital motions
in the Galaxy (used to discuss the population nature of the sample stars)
and the projected rotational velocity as well as the macroturbulent velocity 
(both estimated from the line-broadening width resulting from the spectrum 
fitting analysis). The chemical abundances of various elements are determined
in section 5, followed by section 6 where the characteristics of the metallicity 
distribution and chemical abundances of several key elements are discussed
in connection with other stellar parameters as well as the existence of
planets. In addition, an extra section for discussing the reliability of
[O~{\sc i}] 5577 line as an abundance indicator is prepared as Appendix.

\section{Observational Data}

\subsection{Sample Stars}

The 322 program stars of this study, which are simultaneously the targets 
of Okayama Planet Search Program, were originally selected 
by the following criteria:
\renewcommand{\labelitemi}{---}
\begin{itemize}
\item
Apparently bright ($V < 6$) stars whose declinations are not too low
($\delta > -25^{\circ}$) so as to be effectively observable from Okayama.
\item
The ranges of $B-V$ colors and visual absolute magnitudes are 
within $0.6 < B-V < 1.0$ and $-3 < M_{V} < +2.5$, respectively,
corresponding to the spectral type of late-G giants (G5--K1~III).
\item
Those stars, which are catalogued as apparently variable stars or 
unresolvable binaries, were excluded.
\end{itemize}

The list of these stars is given in table 1, where the HD number, 
the apparent visual magnitude, and the spectral type are presented,
which were taken from the Hipparcos catalogue (ESA 1997). All of 
the 57 stars studied in Paper~I are included in the present sample.
As indicated in the last column of table 1, we regard in this paper 
specific 10 stars (out of 322 objects) as stars hosting planets:
HD~104985 (Sato et al. 2003); 
HD~62509 (Reffert et al. 2006; Hatzes et al. 2006);
HD~28305 (Sato et al. 2007); HD~142091, HD~167042, and HD~210702 
(Johnson et al. 2007; Sato et al., in preparation);
HD~107383\footnote{Strictly speaking, it is not a planet but a brown
dwarf ($m\sin i \sim 20 M_{J}$) which was found in this star. 
However, we included it in this stellar group, which we define to 
be the one hosting substellar companions in a more global sense.} 
(Liu et al. 2008); 
HD~81688, HD~188310, and HD~199665 (Sato et al. 2008).

\subsection{Observations and Data Reductions}

Regarding the basic observational material, we used the ``pure star''
(i.e., without I$_{2}$ cell) spectra covering the $\sim$~5000--6200~$\rm\AA$
region,\footnote{A small fraction of the spectra observed in the early 
time (2000--2001) of the project are of somewhat different spectral ranges 
shifted slightly bluewards (e.g., $\sim$~4800--6000~$\rm\AA$).} 
which were obtained at least once for each star as the standard template 
to be used to derive radial velocity variations by analyzing the spectra 
taken with the I$_{2}$ cell.

Most of the observations were done during the period from 2000 to 2005 
by using the HIDES spectrograph equipped at the coud\'{e} focus of 
the 188 cm reflector at Okayama Astrophysical Observatory.
The reduction of the spectra (bias subtraction, flat-fielding, 
scattered-light subtraction, spectrum extraction, wavelength 
calibration, and continuum normalization) was performed by using 
the ``echelle'' package of the software IRAF\footnote{
IRAF is distributed by the National Optical Astronomy Observatories,
which is operated by the Association of Universities for Research
in Astronomy, Inc. under cooperative agreement with the National 
Science Foundation.} in a standard manner. 
Since 2--3 consecutive frames (mostly 10--20 min exposure for each) 
were observed in a night for each star in many cases, we co-added 
these to improve the signal-to-noise ratio, by which the average S/N 
of most stars turned out to be in the range of $\sim$~100--300. 
The spectral resolving power is $\sim 67000$, corresponding to the standard 
slit width of 200~$\mu$m.

We then determined the stellar radial velocities by comparing these 
spectra with the theoretically synthesized spectra, which were then 
converted into the heliocentric system by using the IRAF task ``dopcor.''
The basic data of our observational material (date of observation, radial
velocities in the laboratory as well as in the heliocentric system) 
are given in the ``obspec.dat'' file in e-table E1.

\section{Fundamental Stellar Parameters}

\subsection{Atmospheric Parameters}

As in our previous studies, we used Kurucz's ATLAS9 grid of 
model atmospheres computed for a wide range of parameters
(Kurucz 1992, 1993), from which atmospheric models for each of 
the stars can be generated by interpolations.

The four atmospheric parameters necessary for constructing model 
atmospheres as well as for abundance determinations [$T_{\rm eff}$ 
(effective temperature), $\log g$ (surface gravity), $v_{\rm t}$ 
(microturbulent velocity dispersion), and [Fe/H] (metallicity, 
represented by the Fe abundance relative to the Sun)] were 
spectroscopically derived from the measured equivalent widths 
($W_{\lambda}$) of Fe~{\sc i} and Fe~{\sc ii} lines based on the 
principle and algorithm described in Takeda, Ohkubo, and Sadakane (2002). 
Practically, we used the TGVIT program (Takeda et al. 2005b) by following 
the same procedure as adopted in  Paper~I (cf. section 3.1 therein).\footnote{
The only difference is that we adopted a more stringent condition for the 
line selection and limited to using lines satisfying $W_{\lambda} \le 120$~m$\rm\AA$
(instead of the upper limit of $150$~m$\rm\AA$ in Paper~I), since it 
revealed that the solutions are rather significantly influenced by 
saturated lines with non-negligible damping wings, where difficulties 
are generally involved in precise $W_{\lambda}$ measurements.}

The finally converged solutions of $T_{\rm eff}$, $\log g$, $v_{\rm t}$,
and [Fe/H] are summarized in table 1. The results are also given in the file
``tgvf\_solution.dat'' in e-table E1, where the intrinsic statistical 
uncertainties (typically $\sim$~10--30K, $\sim$~0.05--0.1~dex, 
0.05--0.1~km~s$^{-1}$, and $\sim$~0.02--0.04~dex) involved in the solutions 
of $T_{\rm eff}$, $\log g$, $v_{\rm t}$, and the Fe abundance estimated 
in the manner described in section 3.2 of Takeda et al. (2002) 
are also presented, though realistic internal errors may be somewhat larger 
than these (especially for $T_{\rm eff}$ and $\log g$; cf. subsection 3.3.).
The measured $W_{\lambda}$ values for each of the adopted Fe~{\sc i} 
and Fe~{\sc ii} lines ($\sim 100$ and $\sim 10$, respectively) and 
the abundances from these lines corresponding to the final 
solutions of the parameters are given for each star in e-table E2
(the results for HD~?????? are contained in the``??????.abd'' file).

\setcounter{figure}{0}
\begin{figure}
  \begin{center}
    \FigureFile(80mm,140mm){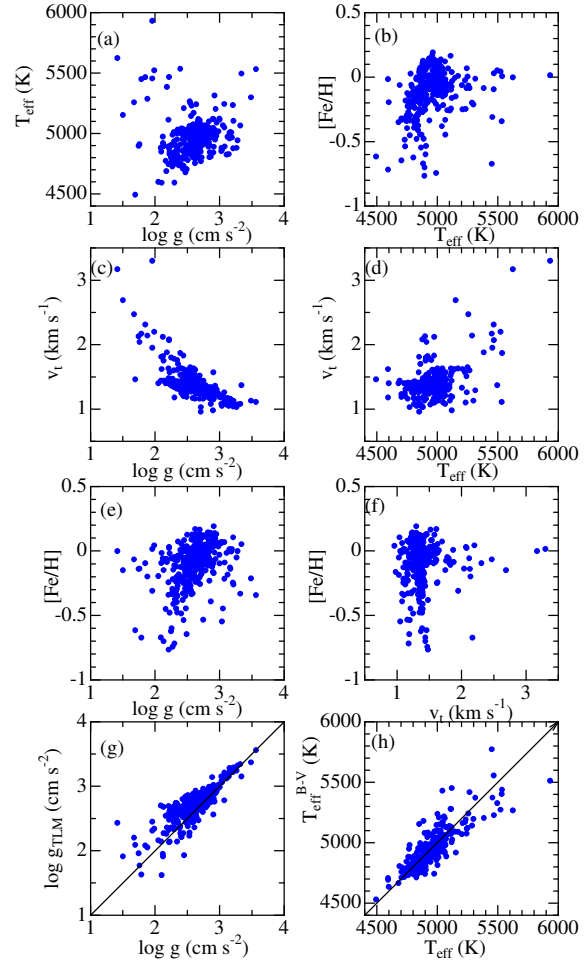}
  \end{center}
\caption{Correlations of the atmospheric parameters
obtained by the spectroscopic method using Fe~{\sc i} and 
Fe~{\sc ii} lines:
(a) $T_{\rm eff}$ vs. $\log g$, (b) [Fe/H] vs. $T_{\rm eff}$,
(c) $v_{\rm t}$ vs. $\log g$, (d) $v_{\rm t}$ vs. $T_{\rm eff}$,
(e) [Fe/H] vs. $\log g$, (f) [Fe/H] vs. $v_{\rm t}$,
(f) $\log g_{\rm TLM}$ (cf. subsection 3.1) vs. $\log g$,
and (g) $T_{\rm eff}^{B-V}$ (cf. subsection 3.1)
vs. $T_{\rm eff}$.
}
\end{figure}

The correlations between any two combinations of these four parameters 
are depicted in figures 1a--f. It is worth noting that different stellar 
groups appear to be involved; i.e., the major population composed of 
many stars having rather similar parameters to each other, 
and the minor population showing more diversed parameter values. 
For example, regarding the $T_{\rm eff}$ vs. $\log g$ relation (figure 1a), 
while $T_{\rm eff}$ tends to be lowered with a decrease in $\log g$ for a 
majority of (densely clumped) stars, there are also stars satisfying 
both low-$\log g$ and high-$T_{\rm eff}$. Actually, the main characteristics 
of our sample stars tend to be determined by the former population 
(red-clump giants), as mentioned in subsection 3.2.
Another remarkable feature is the marked $\log g$-dependence of $v_{\rm t}$
(figure 1c), which clearly indicates the growth of the atmospheric
turbulent velocity field as the surface gravity decreases (an intuitively
reasonable tendency).

As a consistency check of our spectroscopically established $T_{\rm eff}$
and $\log g$, they were compared with $T_{\rm eff}^{B-V}$ (the effective
temperature derived from the $B-V$ color by using the calibration of 
Alonso et al. 1999) and $\log g_{TLM}$ (the surface gravity derived from
$T_{\rm eff}$, $L$, and $M$ determined in subsection 3.2) as shown in 
figures 1g and h, respectively. We may state that no significantly 
systematic discrepancy is seen in these figures, even though the dispersion 
tends to increase toward lower-$\log g$ or higher-$T_{\rm eff}$ stars.

\subsection{Luminosity, Radius, Mass, and Age}

The stellar luminosity ($L$) was derived from the apparent visual
magnitude ($m_{V}$), the parallax ($\pi$) from the Hipparcos catalogue 
(ESA 1997),  the interstellar extinction ($A_{V}$) from 
Arenou et al.'s (1992) table, and the bolometric correction (B.C.) 
from Alonso et al.'s (1999) calibration.\footnote{We used their empirical
formula instead of interpolating Kurucz's (1993) theoretical B.C which we
adopted in Paper~I. As a result, the extent of B.C in this study tends to be 
slightly smaller (by $\sim 0.1$~mag) than that in Paper~I.}
We then obtained the stellar radius ($R$) from $L$ and $T_{\rm eff}$.

\setcounter{figure}{1}
\begin{figure}
  \begin{center}
    \FigureFile(80mm,140mm){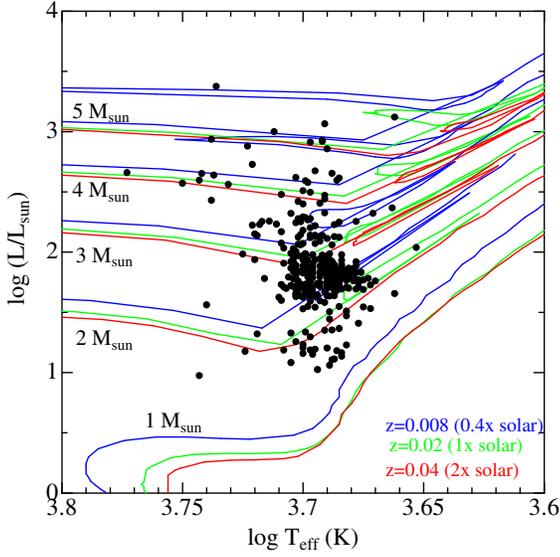}
  \end{center}
\caption{
$\log (L/L_{\odot})$ vs. $\log T_{\rm eff}$ plots based on the data 
given in table 1, along with Lejeune and Schaerer's (2001) 
theoretical evolutionary tracks (for the initial masses of 
1.0, 2.0, 3.0, 4.0, and 5.0 $M_{\odot}$) corresponding to 
three metallicities: $z=0.008$ ([Fe/H] = $-0.4$; blue lines), 
$z=0.02$ ([Fe/H] = $0.0$; green lines), and $z=0.04$ ([Fe/H] = 
$+0.3$; red lines). (Colored only in the electronic edition.)
}
\end{figure}

Now that $T_{\rm eff}$, $L$, and the metallicity 
($z \equiv 0.02\times 10^{\rm [Fe/H]}$, where $z_{\odot}$ is 0.02)
for each star have been established, we can derive the mass ($M$) and 
age ($age$) by comparing the position on the $\log L$ vs. 
$\log T_{\rm eff}$ diagram with Lejeune and Schaerer's (2001) 
theoretical evolutionary tracks 
[$\log L = f_{L}(age|M,z)$, $\log T_{\rm eff} = f_{T}(age|M,z)$], 
as depicted in figure 2. 
The resulting parameter values are presented in table 1, and the
more detailed results including the errors\footnote{
The internal errors in $age$ ($M$) were estimated from the 
difference between $age^{\rm max}$ and $age^{\rm min}$ 
($M^{\rm max}$ and $M^{\rm min}$), which were obtained 
by perturbing the input values of ($\log L$, $\log T_{\rm eff}$, 
and $\log z$) interchangeably by typical amounts of uncertainties
($\Delta \log L$ corresponding to parallax errors given in the Hipparcos 
catalog, $\Delta \log T_{\rm eff}$ of $\pm 0.01$ dex almost corresponding 
to $\sim \pm 100$ K, and $\log z$ of $\pm 0.1$ dex). 
Similarly, the error in $R$ was evaluated from $\Delta \log L$ and 
$\Delta \log T_{\rm eff}$.}
are summarized in the file ``stellar\_params.dat'' in e-table E1.
The inter-relations between such derived $L$, $R$, $age$, $M$ are
displayed in figure 3, where the $M$-dependence of $T_{\rm eff}$, 
$\log g$, and [Fe/H] are also shown. 

\setcounter{figure}{2}
\begin{figure}
  \begin{center}
    \FigureFile(80mm,140mm){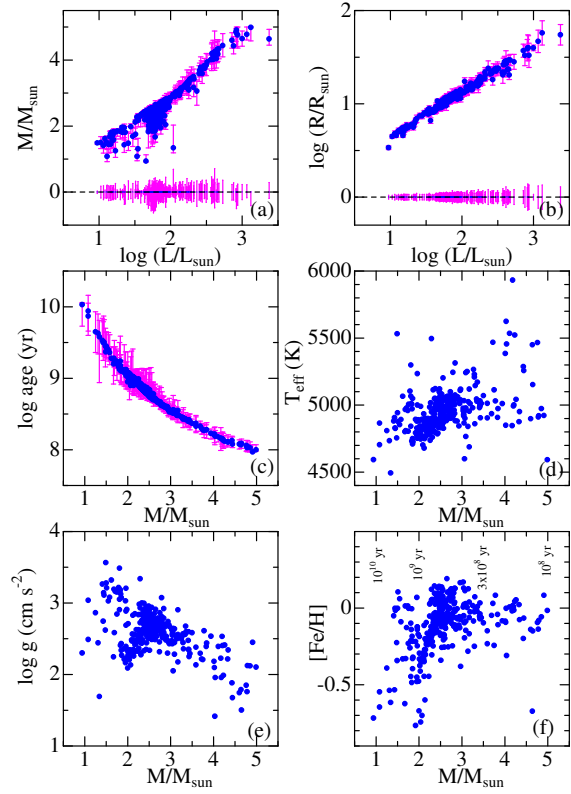}
  \end{center}
\caption{
Correlations between the fundamental stellar parameters
($M$, $L$, $age$) or their dependences upon
atmospheric parameters ($T_{\rm eff}$, $\log g$, and [Fe/H]):
(a) $M$ vs. $\log L$ (vertical ticks indicate the 
internal errors in $M$, which are also replotted in the lower 
part of the figure), (b) $\log R$ (with errors indicated by 
ticks as in panel a) vs. $\log L$, (c) $\log age$ (with errors) 
vs. $M$, (d) $T_{\rm eff}$ vs. $M$, (e) $\log g$ vs. $M$,
and (f) [Fe/H] vs. $M$ (several $age$ values are also indicated).
}
\end{figure}

Several features are recognized from figures 2 and 3:\\
--- As we can clearly see from figure 2, many of our program stars 
clump in the region of $3.67 \ltsim \log T_{\rm eff} \ltsim 3.71$
and $1.6 \ltsim \log L/L_{\odot} \ltsim 2.1$ (corresponding to
$2 \ltsim M/M_{\odot} \ltsim 3$), indicating that 
these objects belong to ``red-clump giants'' (post red-giants 
after the ignition of core He; see Zhao et al. 2001 and the 
references therein).\\
--- Brighter stars tend to be of higher mass almost following 
the relation of $M/M_{\odot} \sim 2 \log (L/L_{\odot}) - 1$,
though stars around $M \sim$ 2--3 $M_{\odot}$ (corresponding to 
red-clump giants) do not necessarily conform to this relation
and show a rather large diversity (figure 3a). \\
--- According to figure 3b, the radius ($R$) is almost a unique function 
of luminosity ($L$) following the relation of $R \propto L^{1/2}$, 
which means that the change in $T_{\rm eff}$ (mostly confined to a 
rather narrow range of several hundred K) does not play any 
significant role here.\\
--- A tight relationship exists between mass ($M$) and age ($age$) as 
$\log age ({\rm yr}) \simeq 10.74 -1.04 (M/M_{\odot}) +0.0999 (M/M_{\odot})^{2}$
(figure 3c). This is reasonably understandable because $age$'s of 
giant stars are practically the same as the main-sequence life time 
(uniquely determined by $M$) which they spent in the past.\\
--- We can see a rough tendency in figure 3d that $T_{\rm eff}$ tends 
to be higher for larger $M$. This may be related to the slope  
of the evolutionary tracks rising toward upper-right (at
$\log T_{\rm eff} \ltsim 3.7$), by which a larger $M$ is assigned to 
a star as its $T_{\rm eff}$ becomes higher (if $L$ remains the same).\\
--- There is a general trend in figure 3e that $\log g$ becomes lower 
toward larger $M$, which is because that the growth rate of 
$(R/R_{\odot})^{2}$ ($\propto L/L_{\odot} \sim 10^{(1+M/M_{\odot})/2}$) with  
increasing $M$ is much larger than that of $M$ itself, though somewhat 
opposite tendency is locally seen for a homogeneous group of red-clump 
giants at $M \sim$ 2--3 $M_{\odot}$ (indicating that $R$ do not 
vary much among these).\\
--- Figure 3f suggests that the metallicity ([Fe/H]) tends to become higher
as $M$ increases, which was also pointed out in Paper~I. This trend
may be interpreted as due to the metallicity dependence of stellar
evolutionary tracks ($L$ tends to be lowered with a decrease in $z$
for a given $M$; cf. figure 2). That is, if a star with a given $L$
is considered, a larger $M$ will be assigned as its metallicity
becomes higher.

\subsection{Comparison with Other Studies}

Figures 4a--f compare the values of $T_{\rm eff}$, $\log g$,
$v_{\rm t}$, [Fe/H], $M$, and $\log R$ derived in this study
with those derived in Paper~I for 57 stars in common.
We may state that both results are almost in agreement 
without any significant systematic differences.
The average [Paper~I $-$ this study] differences ($\pm \sigma$: 
standard deviation) are +28~($\pm 67$)~K, +0.06~($\pm 0.17$)~dex, 
$-0.01$~($\pm 0.04$)~km~s$^{-1}$, +0.05~($\pm 0.06$)~dex,
+0.10~($\pm 0.11$)~$M_{\odot}$, +0.51~($\pm 0.52$)~$R_{\odot}$,
respectively. Since differences in atmospheric parameters are 
essentially due to the changes in the used $W_{\lambda}$ set of
Fe~{\sc i} and Fe~{\sc ii} lines (newly re-measured this time also
for these 57 stars independently from Paper~I; cf. subsection 3.1), 
these results suggest that $\ltsim$~50--100~K, $\ltsim$~0.1--0.2~dex,
$\ltsim$~0.05--0.1~km~s$^{-1}$, and $\ltsim$~0.05--0.1~dex
may be the realistic estimates of internal errors (under
consideration of $W_{\lambda}$-measurement ambiguities) in 
$T_{\rm eff}$, $\log g$, $v_{\rm t}$, and [Fe/H], respectively.

\setcounter{figure}{3}
\begin{figure}
  \begin{center}
    \FigureFile(80mm,140mm){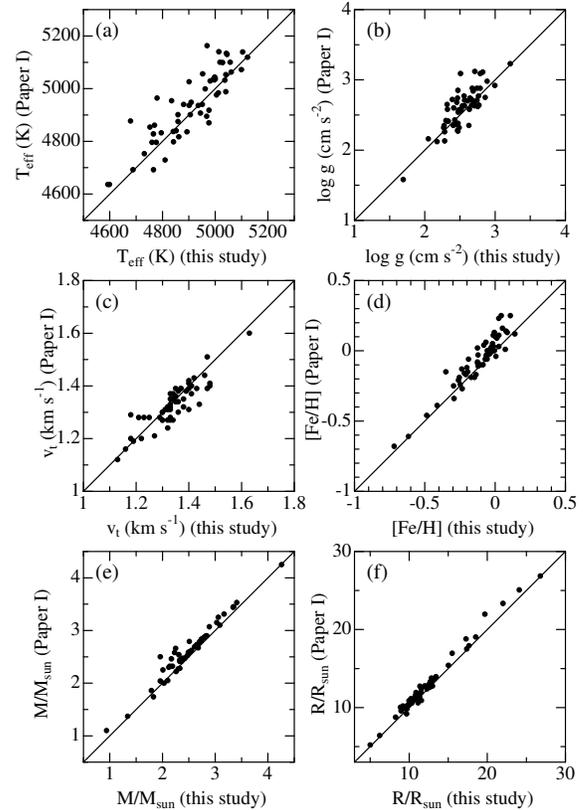}
  \end{center}
\caption{
Comparison of the stellar parameters determined in this study with 
those of 57 stars derived in Paper~I: (a) $T_{\rm eff}$, (b) $\log g$,
(c) $v_{\rm t}$, (d) [Fe/H], (e) $M$, and (f) $R$.
}
\end{figure}

For the purpose of consistency check with the parameter results
of other groups, we refer to McWilliam (1990), which is presumably 
the most extensive investigation so far on 671 G--K giants, as well as 
to the three latest studies available (da Silva et al. 2006, 
Luck \& Heiter 2007, and Hekker \& Mel\'{e}ndez 2007).
How our results are compared with others is graphically displayed
in figures 5, 6, 7, and 8, respectively. 
A glance at these figures suffices to realize that systematic 
differences are more or less observed in many cases; this may simply 
suggest the difficulty of parameter determinations in the case of
giant stars, which critically depends on the method to be adopted
(e.g., photometric vs. spectroscopic etc.) as well as on the data 
to be used (e.g., which lines to be adopted among those of different 
strengths or of atomic parameters). Several remarkable features 
(notable systematic trends specific to our results, considerable 
discrepancies, etc.) are summarized below:\\
--- Our spectroscopically determined $\log g$ values appear to be 
systematically lower by 0.2--0.3~dex (cf. figures 5b, 6b, 7b, and 8b), 
compared to other four previous studies, where
McWilliam (1990) adopted the direct $g$-determination method 
[from $L$, $T_{\rm eff}$, and $M$ (estimated from evolutionary tracks)],
while the spectroscopic method based on Fe~{\sc i} and Fe~{\sc ii} lines 
(similar to that we used) was invoked by da Silva et al. (2006),
Luck and Heiter (2007),\footnote{While Luck and Heiter (2007) published
three different sets of stellar parameters determined in different
ways (``spectroscopic'', ``MARCS75'', and ``physical''; cf. their table 2), 
we used their ``spectroscopic'' parameters for the present 
comparison, which they derived from the Fe~{\sc i} and Fe~{\sc ii} lines
based on the ``new'' MARCS grid of model atmospheres.}, 
and Hekker and Mel\'{e}ndez (2007). Differences in the
used set of lines may have something to do with this tendency,
similarly to the case of $v_{\rm t}$ as mentioned below.\\
--- There is a trend that our $v_{\rm t}$ results are smaller by 
several tenths of km~s$^{-1}$ as compared to others.
This may be attributed to the difference in the lines used
(especially in terms of line strengths), because $v_{\rm t}$ tends to 
be depth-dependent (i.e., increasing with height) in low-gravity stars 
(see, e.g., appendix B in Takeda \& Takada-Hidai 1994).\\
--- Even so, McWilliam's (1990) $v_{\rm t}$ values ($\sim$~2--4~km~s$^{-1}$) 
seem to be exceptionally too large (figure 5b), if we consult the review of 
Gray (1988; see figure 3-8 therein), which indicates that  $v_{\rm t}$ 
generally falls in the range of $\sim$~1--2~km~s$^{-1}$ for G--K giants 
of luminosity class III.\\
--- Luck and Heiter's (2007) $M$ values are appreciably smaller
than our results (figure 7e). We suspect that this may be due to
their use of evolutionary ``isochrones'' (instead of ``tracks'' 
we adopted), since it may cause considerable errors due to
the insufficient time-step of theoretical calculations
when applied to giants under the phase of rapid evolution
(see subsection 3.3 in Paper~I).

\setcounter{figure}{4}
\begin{figure}
  \begin{center}
    \FigureFile(80mm,140mm){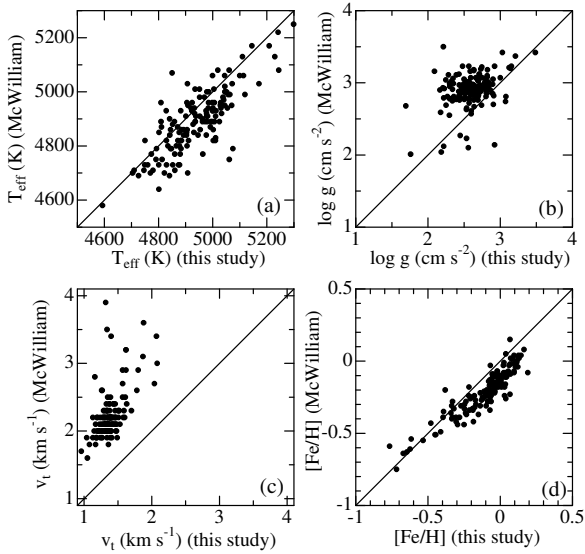}
  \end{center}
\caption{
Comparison of the atmospheric parameters determined in this study
with those derived by McWilliam (1990) for 150 stars in common: 
(a) $T_{\rm eff}$, (b) $\log g$, (c) $v_{\rm t}$, and (d) [Fe/H].
}
\end{figure}

\setcounter{figure}{5}
\begin{figure}
  \begin{center}
    \FigureFile(80mm,140mm){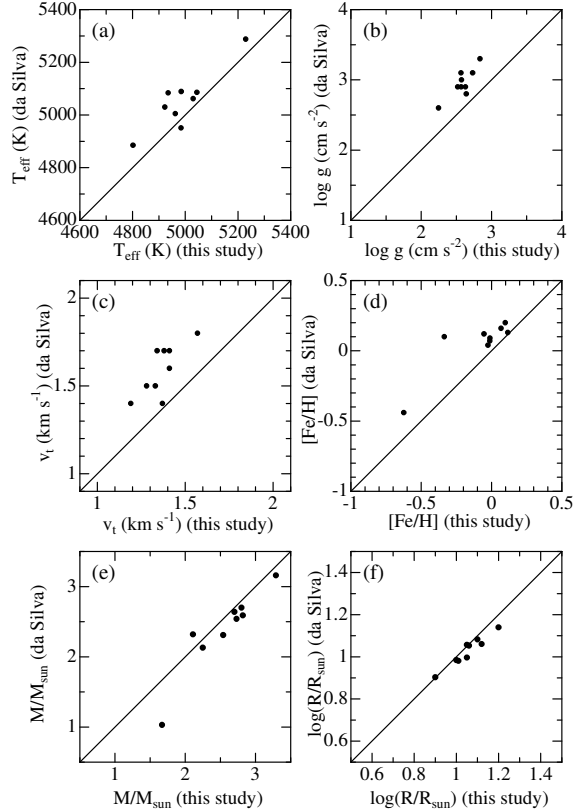}
  \end{center}
\caption{
Comparison of the stellar parameters determined in this study
with those derived by da Silva et al. (2006) for 9 stars in common: 
(a) $T_{\rm eff}$, (b) $\log g$, (c) $v_{\rm t}$, (d) [Fe/H], 
(e) $M$, and (f) $\log R$.
}
\end{figure}

\setcounter{figure}{6}
\begin{figure}
  \begin{center}
    \FigureFile(80mm,140mm){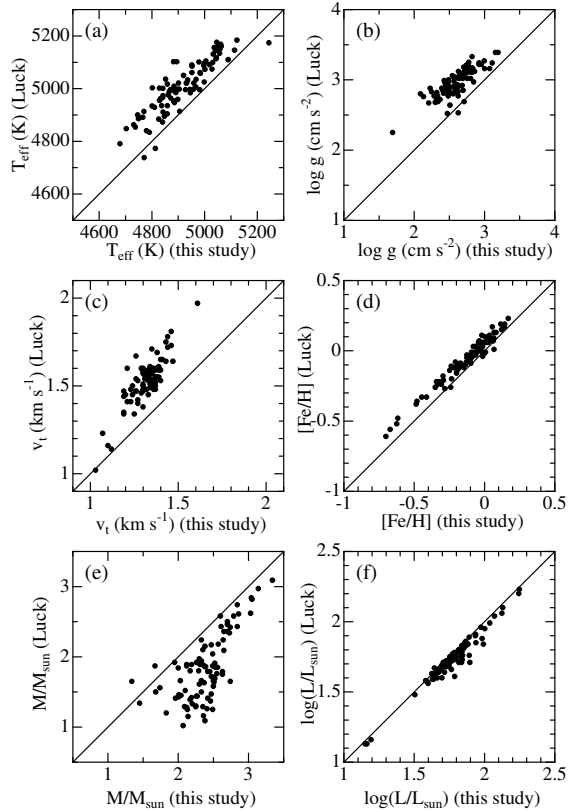}
  \end{center}
\caption{
Comparison of the stellar parameters determined in this study
with those (spectroscopically) derived by Luck and Heiter (2007) 
for 93 stars in common: 
(a) $T_{\rm eff}$, (b) $\log g$, (c) $v_{\rm t}$, (d) [Fe/H], 
(e) $M$, and (f) $\log L$.
}
\end{figure}

\setcounter{figure}{7}
\begin{figure}
  \begin{center}
    \FigureFile(80mm,140mm){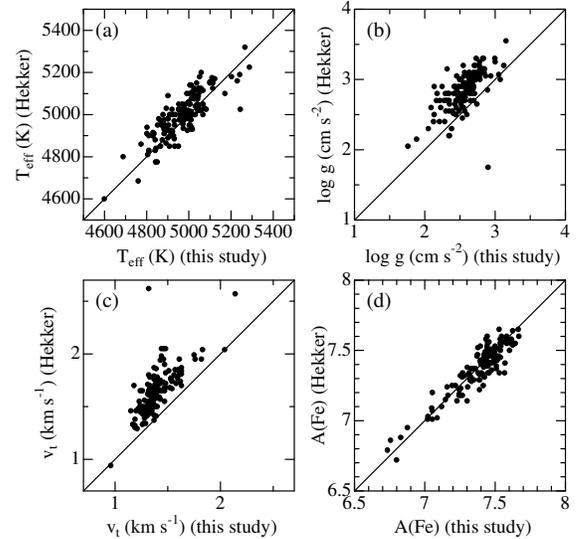}
  \end{center}
\caption{
Comparison of the atmospheric parameters determined in this study
with those derived by Hekker and Mel\'{e}ndez (2007) for 147 stars
in common: (a) $T_{\rm eff}$, (b) $\log g$, (c) $v_{\rm t}$, and 
(d) $A^{\rm Fe}$ (logarithmic Fe abundance in the usual normalization
of $A^{\rm H} = 12.00$).
}
\end{figure}

\section{Kinematics and Stellar Rotation}

\subsection{Kinematic Properties}

In order to examine the kinematic properties of the program stars, 
we computed their orbital motions within the galactic gravitational
potential based on the positional and proper-motion data 
(taken from the Hipparcos catalog) along with the radial-velocity 
data (measured by us), following the procedure described in 
subsection 2.2 of Takeda (2007).
The adopted input data and the resulting solutions of kinematic 
parameters are given the file ``kinepara.dat'' contained in e-table E1.
Figures 9a and b show the correlations of
$z_{\rm max}$ (maximum separation from the galactic plane) vs. 
$V_{\rm LSR}$ (rotation velocity component relative to LSR) and 
$e$ (orbital eccentricity) vs. $\langle R_{\rm g} \rangle$ 
(mean galactocentric radius), respectively. 
Applying Ibukiyama and Arimoto's (2002) classification criteria 
to figure 9a, we can see that most ($\sim 97\%$) stars belong to 
the group of normal thin-disk population, while only 8 stars 
(indicated by open symbols) may be of thick-disk population 
having characteristics of large eccentricity (figure 9b), 
high space-velocity as well as low metallicity (figure 9c), 
and comparatively aged stars of lower-mass (figure 9d).

\setcounter{figure}{8}
\begin{figure}
  \begin{center}
    \FigureFile(80mm,140mm){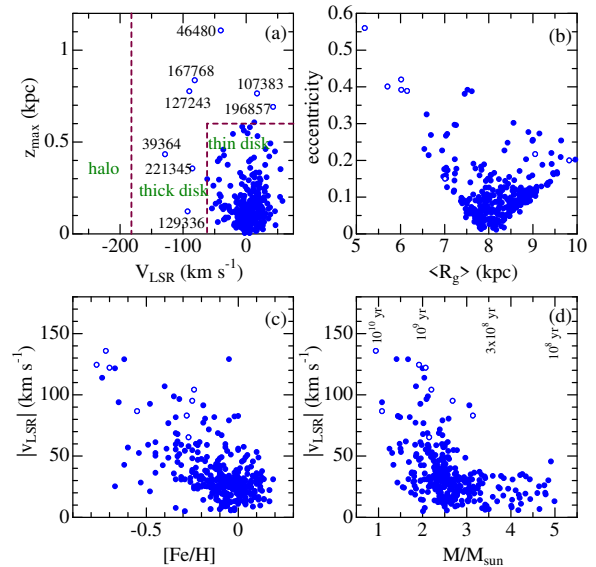}
  \end{center}
\caption{
(a) Correlation diagram between $z_{\rm max}$ (maximum separation 
from the galactic plane) and $V_{\rm LSR}$ (rotation velocity 
component relative to LSR), which may be used for classifying 
the stellar population (the halo/thick-disk/thin-disk boundaries
are also shown by dashed lines according to Ibukiyama \& Arimoto 2002).
Eight stars, which may be thick-disk candidates, are indicated by 
their HD numbers. (b) $e$ (orbital eccentricity) 
plotted against $\langle R_{\rm g} \rangle$ (mean galactocentric radius).
(c) [Fe/H]-dependence of the space velocity relative to LSR
[$|v_{\rm LSR}| \equiv (U_{\rm LSR}^{2} + V_{\rm LSR}^{2} + 
W_{\rm LSR}^{2})^{1/2}$]. (d) $M$-dependence of $|v_{\rm LSR}|$
(approximate $age$'s at four different $M$ values are
also indicated). Shown by open symbols in all four panels are the 
possible thick-disk candidates.
}
\end{figure}

\subsection{Rotational Velocity}

\subsubsection{Modeling of macro-broadening function}

In order to derive the projected rotational velocity ($v_{\rm e}\sin i$) 
from the widths of spectral lines, we made the following assumptions
regarding the line-broadening functions.\\
--- (1) The observed stellar spectrum ($D_{\rm obs}$) is a 
convolution of the modeled intrinsic spectrum ($D_{0}$; computable if 
a model atmosphere, a microturbulence, and elemental abundances are given)
and the total macro-broadening function $f_{\rm M}(v)$; i.e., 
$D_{\rm obs} = D_{0} * f_{\rm M}$ (``*'' means the convolution procedure ).\\
--- (2) The total macrobroadening function is a convolution of
three component functions: the instrumental broadening (denoted as ``ip''), 
rotation (``rt''), and macroturbulence (``mt''); i.e.,
$f_{\rm M} = f_{\rm ip} * f_{\rm rt} * f_{\rm mt}$.\\
--- (3) All of the relevant broadening functions are assumed to 
have the same Gaussian form parameterized by the $e$-folding half-width 
($v_{\alpha}$) as $f_{\alpha}(v) \propto \exp(-v^{2}/v_{\alpha}^{2}$),
where $\alpha$ represents any of the suffixes. Then, 
a simple relation holds between the broadening parameters as
 $v_{\rm M}^{2} = v_{\rm ip}^{2} + v_{\rm rt}^{2} + v_{\rm mt}^{2}$.\\
--- (4) For convenience, we also use the combined broadening function 
$f_{\rm r+m}$, which is the ``macroturbulence + rotation'' function 
defined as $f_{\rm r+m} \equiv f_{\rm rt} * f_{\rm mt}$ (with a relation 
$v_{\rm r+m}^{2} = v_{\rm rt}^{2} + v_{\rm mt}^{2}$). 

\subsubsection{Determination of $v_{r+m}$ from 6080--6089~$\AA$ fitting}

Regarding the actual determination of line broadening for each star, we 
applied the automatic spectrum-fitting technique (Takeda 1995) to the 
6080--6089~$\rm\AA$ region (given the model atmosphere corresponding
to the atmospheric parameters derived in subsection 3.1), which 
successfully establishes such solutions of seven free parameters 
that accomplish the best fit: the abundances of six elements 
(Si, Ti, V, Fe, Co, and Ni) and the total macrobroadening ($v_{\rm M}$). 
See subsection 4.2 of Takeda et al. (2007) for more details. 
Two examples of how the theoretical spectrum corresponding to the final 
solutions matches the observed spectrum are shown in figure 10a. 

Once $v_{\rm M}$ is known, we can obtain $v_{\rm r+m}$ 
($\equiv \sqrt{v_{\rm M}^{2} - v_{\rm ip}^{2}}$ by definition) by
subtracting $v_{\rm ip}$ (2.69~km~s$^{-1}$) corresponding to 
the spectrum resolving power of $R \simeq 67000$.\footnote{Since
the Gaussian FWHM is $3 \times 10^{5} / 67000 \simeq 4.48$~km~s$^{-1}$, 
the corresponding $e$-folding half-width makes 
$4.48/(2\sqrt{\ln 2}) \simeq 2.69$~km~s$^{-1}$.}

\setcounter{figure}{9}
\begin{figure}
  \begin{center}
    \FigureFile(80mm,140mm){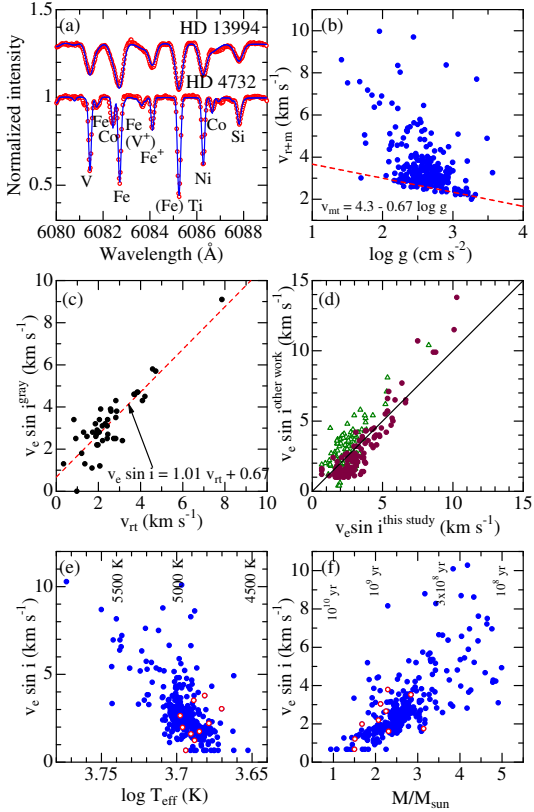}
  \end{center}
\caption{
(a) Examples of the spectrum-synthesis fitting in the 6080--6089~$\rm\AA$ 
region for evaluating the total macro-broadening parameter ($v_{\rm M}$),
from which $v_{\rm r+m}$ (macroscopic broadening velocity field including 
both rotation and macroturbulence) is derived by subtracting the
effect of instrumental broadening. The upper (HD~13994) and lower (HD~4732) 
spectra show the typical cases of higher $v_{\rm e}\sin i$ and lower 
$v_{\rm e}\sin i$, respectively. Identifications of prominent lines are
also given.
(b) Correlation between $v_{\rm r+m}$ and $\log g$, in which we may regard
the lower envelope boundary ($4.3 - 0.67 \log g$; indicated by the dashed line) 
as representing the $\log g$-dependence of $v_{\rm mt}$ (macroturbulence 
velocity dispersion).
(c) Relation between the rotational broadening $v_{\rm rt}$ 
[$\equiv \sqrt{v_{\rm r+m}^{2} - v_{\rm mt}^{2}}$] and the projected
rotational velocity ($v_{\rm e} \sin i$) determined by Gray (1989)
from his elaborate line-profile analysis, plotted for 44 stars
in common. The dashed line shows the linear-regression line (derived
by the least-squares fit), 
$v_{\rm e} \sin i = 1.01 v_{\rm rt} +0.67$, which we adopted to convert
$v_{\rm rt}$ to $v_{\rm e} \sin i$.
(d) Comparison of such derived $v_{\rm e} \sin i$ values with the literature
values: filled circles are those from de Medeiros and Mayor (1999) (128 stars 
in common), while open triangles are those from Massarotti et al. (2008)
(plotted for 96 stars out of 157 stars in common, where 61 stars with 
$v_{\rm e} \sin i^{\rm Massarotti} = 0$ are excluded).
(e) $T_{\rm eff}$-dependence of $v_{\rm e} \sin i$. 10 planet-host stars
are indicated by open circles. (f) $M$-dependence of $v_{\rm e} \sin i$. 
Indicated above are the approximate $age$'s at four different $M$ values, 
while 10 planet-host stars are shown by open circles.
}
\end{figure}

\subsubsection{Separation of rotation and macroturbulence}

Since we now know $v_{\rm r+m}$, the rotational broadening ($v_{\rm rt}$)
can be evaluated by extracting the macroturbulence component ($v_{\rm mt}$) 
from it. Here, we make a practical assumption that ``the macroturbulence
depends only on the surface gravity,'' which we believe to be justified
for the following two reasons.\\
--- (a) According to Gray's (1989) detailed line-profile study on G giants,
we may regard that the $T_{\rm eff}$-dependence of the macroturbulence\footnote{
We can see from figure 7 of Gray (1989) that, while the radial-tangential 
macroturbulence ($\zeta_{\rm RT}$) in late-G giants tends to slightly decrease 
from $\zeta_{\rm RT} \sim 6$~km~s$^{-1}$ at G5~III to 
$\zeta_{\rm RT} \sim 5$~km~s$^{-1}$ at K0~III on the average, this trend 
is not significant compared to the scatter ($\sim 3$~km~s$^{-1}$).}
is almost negligible for our sample stars clustering at the spectral type
of late-G.\\
--- (b) In view of the reasonable connection between the macroturbulence 
and microturbulence (see, e.g., Gray 1988), the remarkably tight 
$\log g$-dependence of $v_{\rm t}$ (increasing with a decrease in $g$;
cf. figure 1c) suggests that the variation of $v_{\rm mt}$ is essentially 
dominated by the change in $\log g$.

The $v_{\rm r+m}$ values are plotted against $\log g$ in figure 10b.
Interestingly, we can recognize in this figure a clear-cut boundary line
($v_{\rm r+m}^{\rm boundary} \simeq 4.3 - 0.67 \log g$), below which
no stars are seen. Considering that the contribution of projected rotational
velocity can be as small as zero (in case of nearly pole-on stars), we can 
reasonably assume that this lower boundary represents the case of 
$v_{\rm rt} \simeq 0$, which leads to the relation we use for estimating 
the macroturbulence
\begin{equation}
v_{\rm mt} = 4.3 - 0.67 \log g.
\end{equation}
We point out that the $v_{\rm mt}$ range of $\sim$~2--3~km~s$^{-1}$
derived from this equation is just consistent with Gray's (1989)
result of $\zeta_{\rm RT} \sim$~5--6~km~s$^{-1}$, since 
the relationship of $v_{\rm mt} \simeq 0.4 \zeta_{\rm RT}$
is expected to hold.\footnote{While the $v^{*}$ value corresponding to
the half-maximum is $0.83 v_{\rm mt} (=\sqrt{\ln 2} \, v_{\rm mt})$ for the 
Gaussian macroturbulence function, it is $v^{*} \simeq 0.35 \zeta_{\rm RT}$
for the case of the radial-tangential-type macroturbulence function
(see, e,g., figure 17.5 in Gray 2005).
That is, on the requirement that the FWHM of two broadening functions of
different types be equal, we obtain 
$v_{\rm mt} \simeq (0.35/0.83) \zeta_{\rm RT} \simeq 0.42 \zeta_{\rm RT}$.
Quite similarly, since $v^{*} \simeq 0.78 v_{\rm e}\sin i$ for the realistic 
rotational broadening function (e,g., figure 18.5 in Gray 2005), we have 
$v_{\rm rt} \simeq (0.78/0.83) v_{\rm e}\sin i \simeq 0.94 v_{\rm e}\sin i$
as the relation between $v_{\rm rt}$ and $v_{\rm e}\sin i$.}

\subsubsection{Calibration of $v_{e} sin i$}

Now that the macroturbulence ($v_{\rm mt}$) for each star has been 
assigned, we can obtain the rotational broadening parameter ($v_{\rm rt}$) 
from the already known $v_{\rm r+m}$ as
$v_{\rm rt} = \sqrt{v_{\rm r+m}^{2} - v_{\rm mt}^{2}}$.
However, since our modeling is based on a rather rough approximation
of Gaussian rotational broadening, we have to find an appropriate
calibration relation connecting $v_{\rm rt}$ and $v_{\rm e}\sin i$, 
for which we invoke Gray's (1989) $v_{\rm e} \sin i$ results for 
G giants derived from his elaborate line-profile analysis.
Figure 10c shows the correlation of our $v_{\rm rt}$ and Gray's (1989)
$v_{\rm e}\sin i$ for 44 stars in common. 
We then have a linear-regression relation, 
\begin{equation}
v_{\rm e} \sin i = 1.01 v_{\rm rt} + 0.67,
\end{equation}
which we finally adopted to obtain $v_{\rm e}\sin i$.
We point out that this proportionality factor of 1.01 is quite reasonable,
considering the value of 1.05 (=1/0.94) expected from a rough estimation 
(see footnote 11).

The resulting values of $v_{\rm e}\sin i$ (along with $v_{\rm r+m}$ and 
$v_{\rm mt}$) are given in the file ``profit6085.dat'' contained 
in e-table E1, where the abundances of Si, Ti, V, Fe, Co, and Ni 
(derived as by-products of 6080--6089~$\rm\AA$ fitting) are also presented.
As shown in figure 10d, our $v_{\rm e}\sin i$ results are in reasonable agreement 
(even though ours tend to be slightly smaller at the high $v_{\rm e}\sin i$ 
range of $\sim 10$~km~s$^{-1}$) with the recent two extensive determinations 
by de Medeiros and Mayor (1999) and Massarotti et al. (2008).
Note that, although these two studies are based on different techniques 
(cross-correlation method with CORAVEL, line-broadening width measurement 
similar to ours), they both used Gray's results as the calibration standards.

Figures 10e and f display the correlations of $v_{\rm e}\sin i$ vs. 
$T_{\rm eff}$ and $v_{\rm e}\sin i$ vs. $M$, respectively.
We can confirm in figure 10e an apparent rotational break at 
$T_{\rm eff} \sim 5000$~K, below which $v_{\rm e} \sin i$ quickly 
falls off, consistently with the conclusion of Gray (1989).
The tendency of increasing $v_{\rm e}$ toward larger $M$ (figure 10f)
may be interpreted as mainly due to the positive correlation between
$M$ and $T_{\rm eff}$ (cf. figure 3d), though it may partly reflect
the real $M$-dependence of the angular momentum.
Since the distribution of $v_{\rm e}\sin i$ for 10 planet-host stars
does not differ much from that of non-planet-host stars (figures 10e and f), 
we could not nominate any clear such candidates that have acquired excess 
angular momentum by ingestion of planets (see also Massarotti et al. 2008).

\section{Elemental Abundances}

The abundances of 17 elements (C, O, Na, Si, Ca, Sc, Ti, V, Cr, Mn, Co, 
Ni, Cu, Y, Ce, Pr, Nd) relative to the Sun were derived from the measured 
equivalent widths in the same way as described in subsection 4.1 of 
Paper~I,\footnote{One difference is that (unlike Paper~I) we did not 
determine the abundances of elements with $Z > 60$ (e.g., Gd, Hf) this time, 
because they are based mostly on only one line and thus unreliable.} 
which should be consulted for more details. 

The detailed line-by-line results of relative-to-Sun differential 
abundances ($\Delta$) and their average ([X/H] $\equiv \langle \Delta \rangle$) 
are presented in e-table E3 (the results for HD~?????? are contained in 
the``??????.cmb'' file). Also, the [X/H] values for each of the species 
are summarized in the file ``xhresults.dat'' of e-table E1.
The [X/Fe] ratios ($\equiv$ [X/H] $-$ [Fe/H]) 
are plotted against [Fe/H] in figure 11, where the results corresponding to
the abundances (of Si, Ti, V, Fe, Co, and Ni) derived from 6080--6089~$\rm\AA$
fitting are also shown for comparison. We can see by comparing this figure
with figure 7 of Paper~I that the characteristic trend of [X/Fe] vs. [Fe/H]
exhibited by each species (useful for discussing the chemical
evolution in the Galaxy) has become more manifest in the present study 
thanks to the increased number of stars. 


\section{Abundance and Metallicity Characteristics}

Now that we have accomplished our main purpose of determining
the parameters and surface abundances of 322 late-G giants 
in the preceding sections 2--5, some discussion based on these results 
may be appropriate here regarding the notable features seen in the derived 
abundances and the metallicity, especially in connection with their 
dependence upon stellar parameters or with the nature of planet-host stars.

\subsection{Abundance Anomalies in C, O, and Na}

By comparing figure 11 with Takeda's (2007) figure 12, we can confirm 
that the behaviors of [X/Fe] vs. [Fe/H] plots for these late-G giants are 
mostly similar to those of F--G--K dwarfs in the solar neighborhood
for many comparatively heavier species (i.e., Si, Ca, Sc, Ti, V, Cr,
Mn, Co, Ni, Cu), which suggests that the abundance trends of these elements
are reasonably understood as due to the chemical evolution of the Galaxy.

However, the situation is different for the three lighter elements 
(C, O, Na), as can be recognized when figures 11a, b, c are compared
with Takeda and Honda's (2005) figures 6a and 6c and Takeda's (2007)
figure 12a, respectively. Namely, the zero point (the value of [X/Fe]
corresponding to the solar metallicity of [Fe/H] = 0) is appreciably 
discrepant from zero ([C/Fe]~$<$~0, [O/Fe]~$<$~0,\footnote{Unfortunately,
we are not confident with the O abundance derived from only one forbidden 
[O~{\sc i}] line at 5577.34~$\rm\AA$. Actually, it is probable
that our [O/H] values derived for these giant stars are significantly 
underestimated by as much as 0.3--0.4~dex (the zero point might have 
to be shifted down to the position shown by the dotted line in figure 11b). 
This problem is separately discussed in Appendix more in detail.
Anyway, we use in this discussion our [O/H] values as they are, hoping that 
they still correctly describe the {\it relative} behaviors 
(i.e., the slope of $|$[O/Fe]/[Fe/H]$|$ or $|$[O/Fe]/[C/Fe]$|$, for example) 
even if considerable zero-point errors are involved in [O/H] or [O/Fe] in 
the absolute sense.} and [Na/Fe]~$>$~0) and the slope of 
$|$[X/Fe]/[Fe/H]$|\sim 1$ for C and O is appreciably steeper than the case 
of dwarf stars ($\sim$~0.2 for C and $\sim$~0.4 for O; cf. subsection 5.1 
in Takeda \& Honda 2005).
More interestingly, we can observe a correlation between C and O and 
an anti-correlation between C and Na (and also between O and Na) as 
shown in figures 12a, b, and c. Besides, these C, O, and Na abundances 
appear to depend upon the stellar mass (figures 12e, f, and g).

It is then natural to consider that the abundances of these three elements 
(C, O, Na) in the photosphere of late-G giants have suffered appreciable 
changes (a decrease in C and O, an increase in Na) from their original 
composition and the effect of such ``a posteriori'' abundance changes 
becomes progressively pronounced as $M$ becomes larger. Regarding the 
mechanism for this cause, it is likely to be mixing of the H-burning product 
dredged-up from the deep interior, where C and O are reduced by the CN- and 
ON-cycles while Na is enriched by the NeNa-cycle (as already speculated 
in subsection 5.3 of Paper~I).
Further, the extent of this mixing-induced anomaly tends to be larger for 
higher-metallicity stars because of the positive correlation between [Fe/H] 
and $M$ (cf. figure 3f), which reasonably accounts for the trends in 
[C/Fe] vs. [Fe/H] and [O/Fe] vs. [Fe/H] (steep gradient) as well as in 
[Na/Fe] vs. [Fe/H] (conspicuous raise toward [Fe/H] $\gtsim 0$) seen in 
figures 11a, b, and c.

This scenario naturally explains the relationship between [C/Si] and [Si/H]
(cf. subsection 5.2 of Paper~I\footnote{In that paper this tendency was 
discussed in terms of the selective depletion of refractory elements
(such as Si) while the volatile species (such as C) remain unchanged,
which is seen in $\lambda$ Boo-type stars. Instead, we now consider 
it is C that has acquired anomaly.}); that is, since the reduction of 
photospheric C  becomes more efficient at higher metallicity 
as well as higher mass, the tendency of anticorrelation seen in [C/Si] vs. 
[Si/H] (figure 12d) and [C/Si] vs. $M$ (figure 12h) is reasonably understood.

We remark, however, that such abundance changes due to evolution-induced
envelope mixing in late-G giants of 1.5--5 $M_{\odot}$ has not yet been 
theoretically justified at least for O and Na. Namely, according to
the canonical stellar evolution calculations (e.g.; Lejeune \& Schaerer 2001), 
such giant stars of intermediate-mass only show a sign of CN-cycled products 
(C-deficient and N-enriched material, while O and Na are essentially unchanged) 
because the mixing is not so deep as to salvage ON-cycle or NeNa-cycle products.
Therefore, it would be highly desirable to investigate from the theoretical side
whether such an O-deficiency and Na-enrichment is ever feasible or not in the 
photosphere of late-G giants (including red-clump giants), such as seen in
old globular cluster stars (where Na vs. O anti-correlation is reported;
see, e.g., Kraft 1994) or high-mass supergiants (Na is generally overabundant as 
discussed in Takeda \& Takada-Hidai 1994; and the possibility of O-deficiency 
due to mixing of ON-cycled gas was suspected by Luck \& Lambert 1985).

Also, we would again call attention to the poor reliability (at least 
in the absolute sense) of our O-abundances derived from the [O~{\sc i}] 5577 
line (cf. footnote 13 and the appendix), which may contain considerable 
zero-point error. 
Hence, as far as the results involving [O/Fe] or [O/H] are concerned, further 
check or examinations using various other lines (e.g., [O~{\sc i}] 6300/6363 
or O~{\sc i} 7771--5 triplet) would be required before reaching the final 
conclusion.

\setcounter{figure}{11}
\begin{figure}
  \begin{center}
    \FigureFile(80mm,140mm){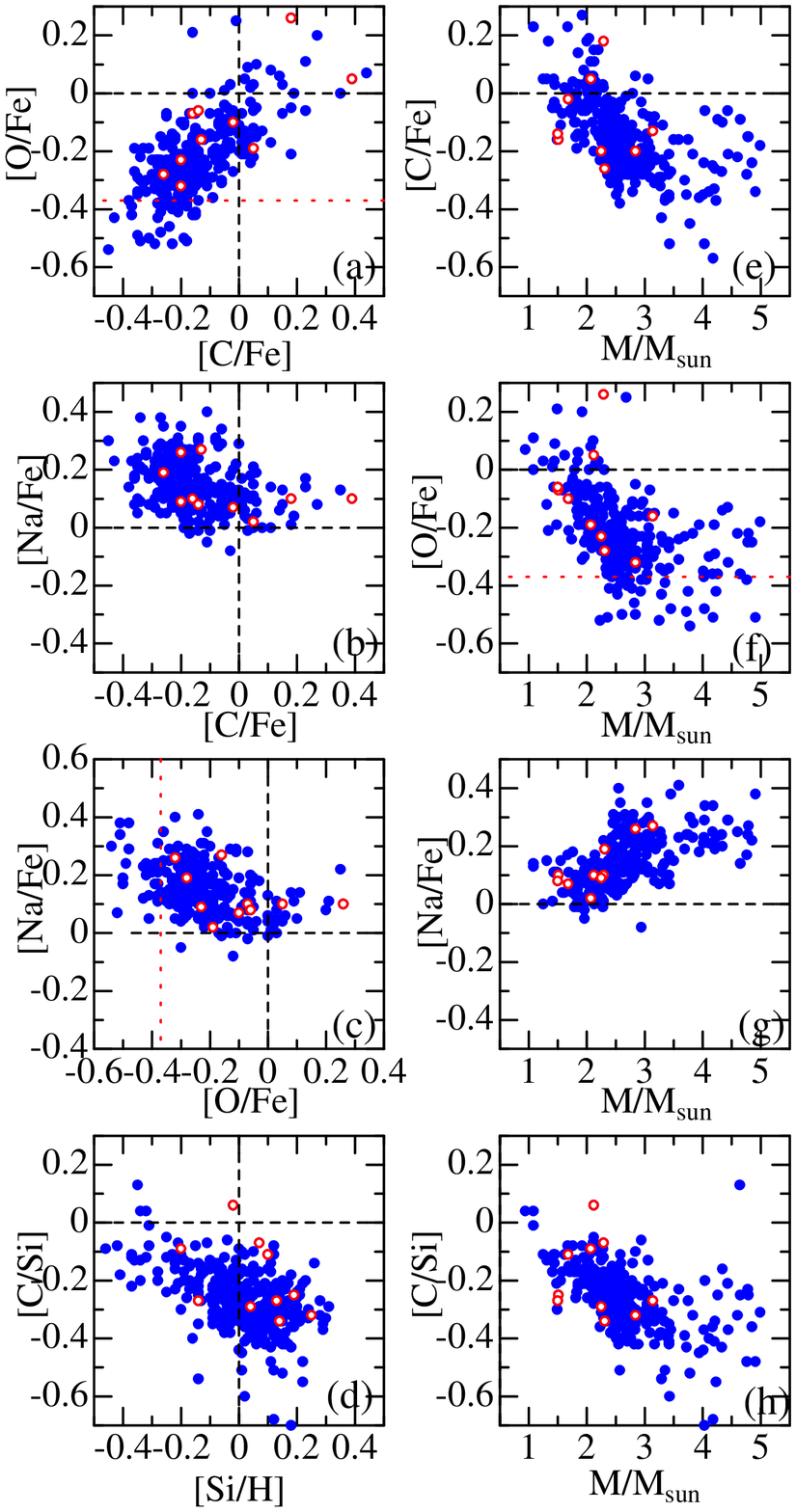}
  \end{center}
\caption{
Correlations between the abundance ratios and their $M$-dependences:
(a) [O/Fe] vs. [C/Fe], (b) [Na/Fe] vs. [C/Fe], (c) [Na/Fe] vs. [O/Fe],
(d) [C/Si] vs. [Si/H], (e) [C/Fe] vs. $M$, (f) [O/Fe] vs. $M$,
(g) [Na/Fe] vs. $M$, and (h) [C/Si] vs. $M$. In panels (a), (c), and 
(f), the position of [O/Fe] = $-0.37$ is shown by a dotted line,
to which the zero-point of [O/Fe] might be lowered (cf. Appendix). 
Planet-host stars are indicated by open symbols.
}
\end{figure}

\subsection{Metallicity Distribution}

Form the viewpoint of planet formation, an important subject is to examine 
whether the metallicity distribution of planet-harboring giants shows 
any difference from that of ordinary giants without planets. 
Figures 13a and b show the histogram of [Fe/H] distribution for our 322 
targets, separated for 312 non-planet-host stars (non-PHS) and 
10 planet-host stars (PHS), respectively. We can recognize in figure 13a 
that [Fe/H] has a characteristic distribution, which is peaked at a 
slightly subsolar value ($\sim -0.1$) with a gradual/steep decline 
toward lower/higher metallicity. It is also worth noting that no 
super-metal-rich stars ([Fe/H] $> +0.2$) are found in our sample. 
Regarding the [Fe/H] trend of planet-host giants, although the number of 
the sample is too small to make any definite argument, the metallicity range 
(from $\sim -0.4$ to $\sim +0.2$ centering around $\sim -0.1$) is quite 
similar (figure 13b) to that of non-planet-host stars; 
actually, the average values are almost indistinguishable 
($\langle$[Fe/H]$\rangle^{\rm nonPHS} = -0.11$
and $\langle$[Fe/H]$\rangle^{\rm PHS} = -0.13$).
Hence, we conclude that there is no essential difference in the metallicity
distribution between planet-host giants and non-planet-host giants, 
which makes marked contrast to the case of F--G--K dwarfs,\footnote{
In addition, we note from figures 11 and 12 that this argument 
also holds for the relative abundance patterns (i.e.,, distribution of 
[X/Fe] ratios), which means that planet-host giants and non-planet-host
giants are practically indiscernible in terms of the chemical abundance 
properties in general.} 
where plane-harboring stars tend to be generally metal-rich (see, e.g., 
Gonzalez 2003 or Udry \& Santos 2007, and the references therein).

\setcounter{figure}{12}
\begin{figure}
  \begin{center}
    \FigureFile(80mm,140mm){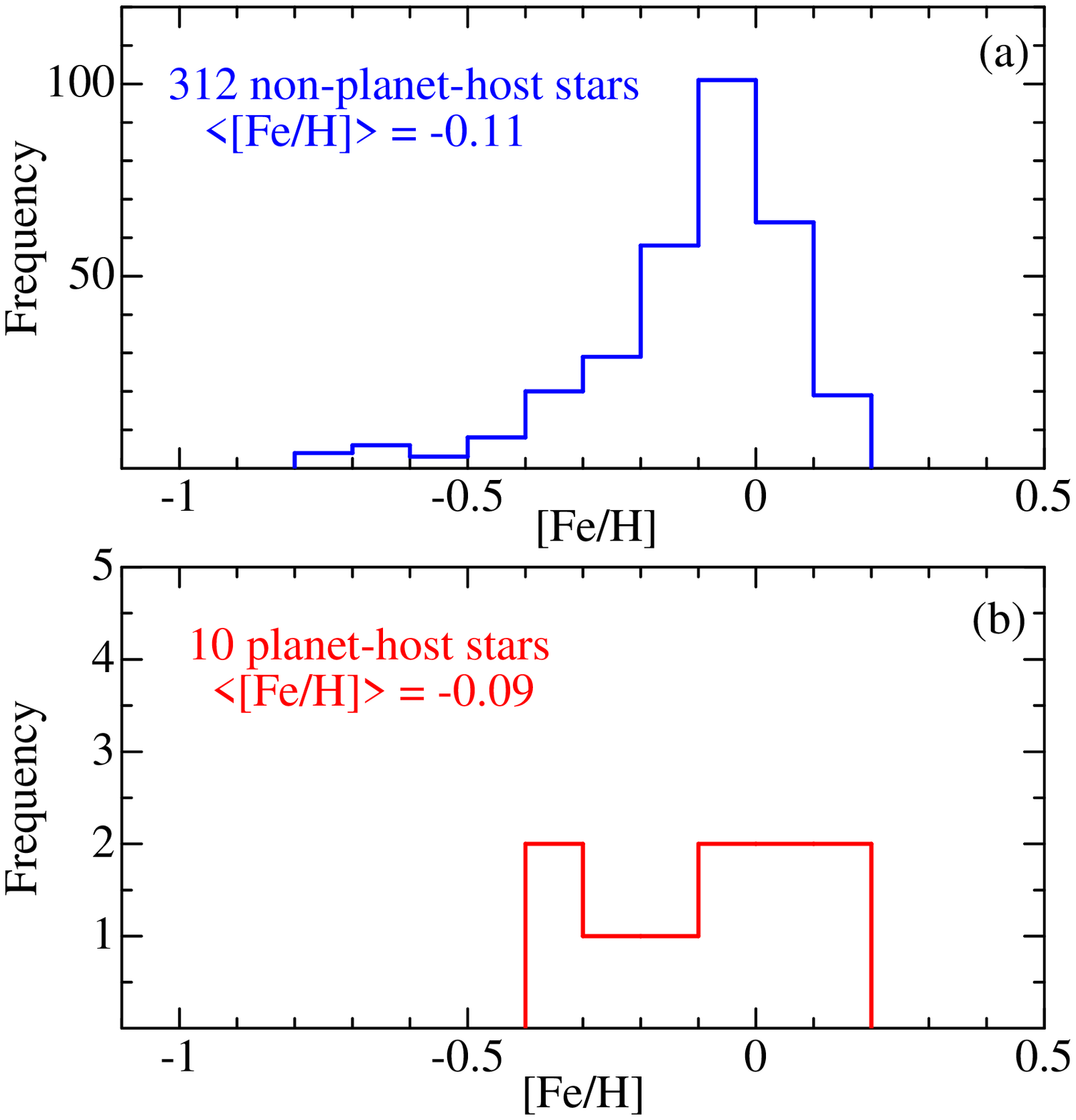}
  \end{center}
\caption{
Metallicity distribution (histogram of the numbers of stars per 
0.1 dex bin in [Fe/H]) of the program stars, separately shown for 
(a) 312 non-planet-host stars and (b) 10 planet-host stars.
}
\end{figure}

This consequence (lack of metal-rich tendency in planet-host giants) is 
in fair agreement with the result of Pasquini et al. (2007). However, we can 
not lend support for Hekker and Mel\'{e}ndez's (2007) contradictory argument
that planet-host giants are more metal-rich by 0.13 dex as compared to
the large sample of ordinary giants. We suspect that this may reflect
their sample choice of planet-host stars, in which not so much giants as 
subgiant stars of near-solar-mass are included (i.e., the general trend may 
be partly affected/contaminated by the characteristics of higher-gravity stars). 
We would further point out that they used ``literature [Fe/H] values'' taken 
from various sources for planet-host giants, which were compared with their 
own [Fe/H] results of normal giants; this makes us feel that their results 
had better be viewed with caution, especially when delicate abundance 
differences as small as $\sim 0.1$~dex are involved. 

How should we interpret the absence of metal-rich trend in planet-host giants 
in contrast to the case of dwarfs (i.e., the fact that planets can form around 
rather metal-poor intermediate-mass stars)?  Some explanations may be 
possible even within the framework of the standard core-accretion theory 
favoring the metal-rich condition (e.g., Hayashi et al. 1985; Ida \& Lin 2004):\\
--- (1) Since the mass of the proto-planetary disk tends to be generally large 
for massive stars, sufficient material for core-formation may still be
available even in the metal-deficient condition as low as [Fe/H] $\sim -0.3$,
making the planet formation feasible.\\
--- (2) While planet-formation may proceed efficiently in the metal-rich case, 
such planets (formed in a rather short time scale) are apt to migrate inward 
(because substantial amount of disk-gas may still remain without being 
dissipated), which would not survive as planets around giants because of 
being engulfed. In this sense, not-so-metal-rich system might be even more 
favorable for planet-detection around giants.\\
--- (3) Alternatively, the planet-formation around intermediate-mass stars 
might occur by a mechanism other than the canonical core-accretion. 
As a matter of fact, in order to explain the metal-poor long tail of 
[Fe/H] distribution in planet-host dwarfs (e.g., Udry \& Santos 2007), 
it has been argued that two different planet-formation processes may be 
coexistent; i.e., the metallicity-dependent core-accretion process, 
and the disk-instability mechanism (e.g., Boss 2002) which is considered 
to take place almost independently on the metallicity. If planets around 
intermediate stars are preferentially formed by the latter mechanism, 
the observational fact may be reasonably explained.

\setcounter{figure}{13}
\begin{figure}
  \begin{center}
    \FigureFile(80mm,140mm){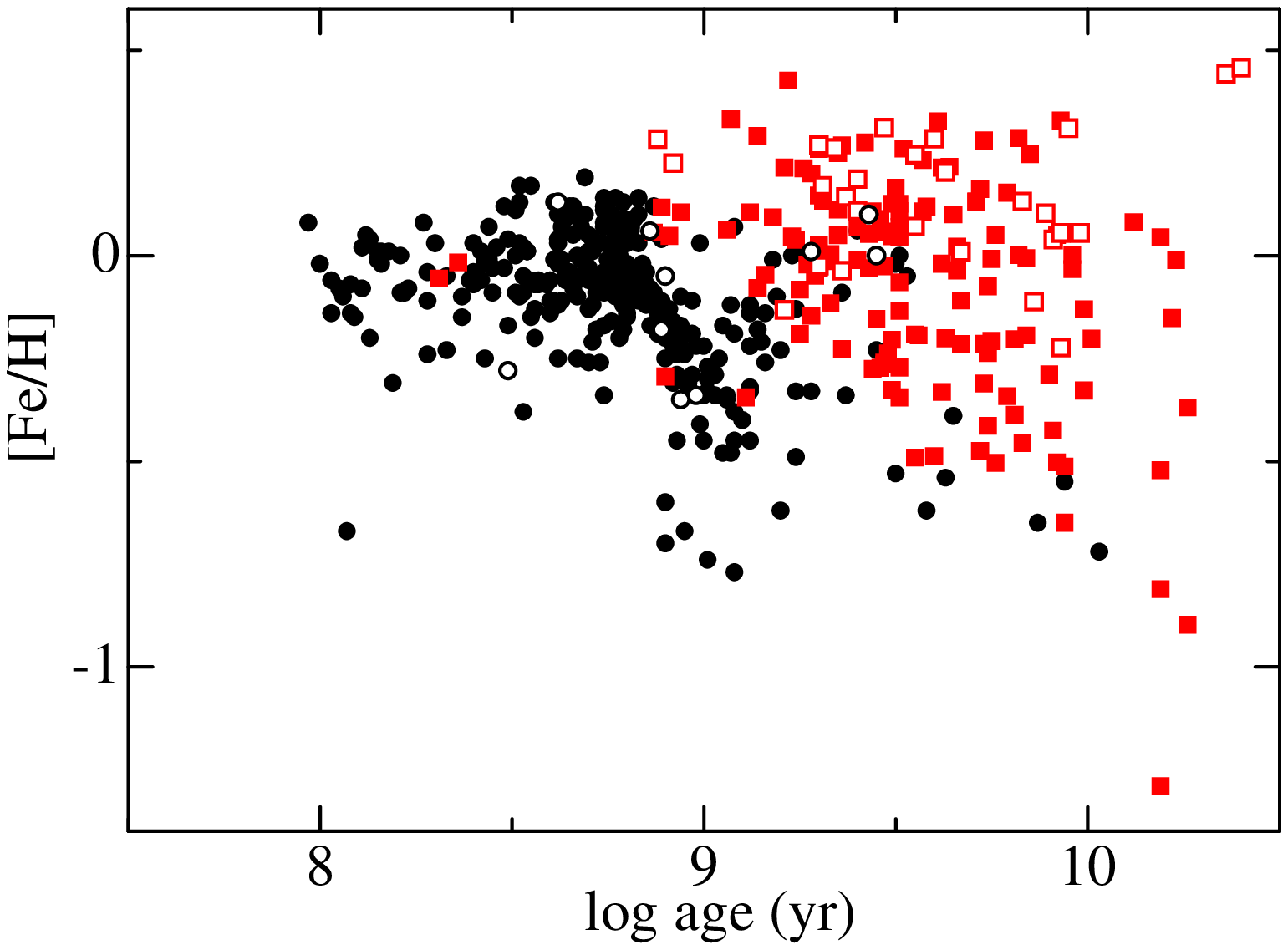}
  \end{center}
\caption{
Age--metallicity relation. Circles show the results for
322 late-G giants investigated in this study, while
squares are those for the 160 F--G--K dwarfs
taken from Takeda (2007). Planet-host stars (10 out of 322
for the former and 27 out of 160 for the latter) are 
indicated by open symbols.
}
\end{figure}

Finally, the age--metallicity relation for 322 late-G giants is displayed
in figure 14, where the similar relation obtained for 160 F--G--K dwarfs
taken from Takeda (2007) is also shown for comparison.
As discussed above, the metal-rich tendency of planet-host dwarfs and 
the absence of such trend for planet-host giants are manifestly observed.
We can also see that planet-host giants so far reported have ages older 
than several $\times 10^{8}$ years (corresponding to $M \ltsim 3 M_{\odot}$), 
which might be related to the time scale of planet formation. 
A strange feature recognized from this figure is the apparent discontinuity
in the metallicity distribution between giants and dwarfs; that is,
while the large scatter of [Fe/H]$_{\rm dwarfs}$ seen in old stars
($age \sim 10^{10}$~yr) tends to converge toward medium-aged stars 
($age \sim 10^{9}$~yr), a large spread reappears in [Fe/H]$_{\rm giants}$
at $age \sim 10^{9}$~yr which again shrinks toward young stars 
($age \sim 10^{8}$~yr). Also, the metallicity upper-limit of giants 
($\sim +0.2$~dex) is lower than that for dwarfs ($\sim +0.4$~dex),
which results in the ``lack of super-metal-rich giants'' as already 
remarked at the beginning of this subsection. If this trend is real, 
it might serve as a clue to investigate the past history of galactic 
chemical evolution (e.g., a special event such as a substantial infall 
of metal-poor primordial gas might have happened $\sim 10^{9}$ years ago).
It would thus be desirable/necessary to check on late B through F 
main-sequence stars (i.e., progenitors of giants) whether the same 
tendency as seen in these G-giants is observed.

\section{Summary and Conclusion}

For the purpose of clarifying the properties of the targets of 
Okayama Planet Search Program, we conducted a comprehensive investigation 
of stellar parameters and photospheric chemical abundances for 322 
intermediate-mass late-G giants (including 10 planet-host stars).

The atmospheric parameters ($T_{\rm eff}$, $\log g$, $v_{\rm t}$, and [Fe/H])
were determined from the equivalent widths of Fe~{\sc i} and Fe~{\sc ii} lines,
and the mass and age were estimated from the position on the HR diagram
with the help of stellar evolutionary tracks. Many of our program stars 
were found to be ``red-clump giants.''

The kinematic parameters ($z_{\rm max}$, $V_{\rm LSR}$, $e$, 
$\langle R_{\rm g} \rangle$, etc.) were evaluated by computing the
orbital motion in a given galactic gravitational potential.
Most stars ($\sim 97\%$) appear to belong to the thin-disk population,
though eight stars are suspected to be of thick-disk origin.

The projected rotational velocities ($v_{\rm e}\sin i$) were determined
from the width of macro-broadening function, evaluated by the spectrum-fitting
in the 6080--6089~$\rm\AA$ region, by subtracting the effect of macroturbulence.
We confirmed a rotational break at $T_{\rm eff} \sim 5000$~K, below which
$v_{\rm e}\sin i$ quickly falls off.

The photospheric chemical abundances (differential values relative to the Sun) 
of 17 elements (C, O, Na, Si, Ca, Sc, Ti, V, Cr, Mn, Co, Ni, Cu, Y, Ce, Pr, Nd) 
were derived from the equivalent widths of selected spectral lines.
The resulting [X/Fe] vs. [Fe/H] relations for giants were found to be similar 
to those of F--G--K dwarfs for most of the heavier elements (Si--Cu),
indicating that the abundance trends of these elements may be understood 
within the framework of the galactic chemical evolution.

However, abundance peculiarities were found in C, O, and Na, in the sense 
that C and O are deficient while Na is enriched, and the extents of
these anomalies appear to increase with the stellar mass. We thus suspect
that the surface abundances of these elements have suffered changes caused 
by mixing of H-burning products (CN-, ON-, and NeNa-cycle) salvaged from 
the deep interior, though our results for O derived from the [O~{\sc i}] 5577 
line should be regarded with caution which may be considerably underestimated.

The metallicity distribution of planet-host giants was found to be 
almost the same as that of non-planet-host giants (i.e., planets are equally 
found for metal-poor as well as metal-rich giants), which makes marked
contrast to the case of planet-host dwarfs tending to be metal-rich. 
Any theory for planet-formation around intermediate-mass should
account for this fact.

When the metallicities of these comparatively young (typical $age$ of 
$\sim 10^{9}$~yr) giants are compared with those of F--G--K dwarfs
(mainly $10^{8}$~yr~$\ltsim age \ltsim 10^{9}$~yr), a discontinuity
appears to exist between these two groups, and [Fe/H]$_{\rm giants}$
tend to be somewhat lower than [Fe/H]$_{\rm giants}$ at the same age 
with an apparent lack of super-metal-rich ([Fe/H] $>$ 0.2) giants.
\newline
\newline
This study is based on the observational material which has been 
accumulated during the course of the Okayama Planet Search Program 
over the past 7 years. We are grateful to all the project members 
for their collaboration and encouragement, as well as to the observatory 
staff for their helpful support in the observations.
Special thanks are due to M. Omiya, E. Toyota, S. Masuda, E. Kambe, 
and H. Izumiura, who have made particularly large contributions
in carrying out the observations. We further thank S. Ida 
for his insightful comments from the theoretical side concerning the 
metallicity-independence of planet-host giants.
Financial supports by Grant-in-Aid for Young Scientists (B) No.17740106
(to B.S.) and by ``The 21st Century COE Program: The Origin and Evolution
of Planetary Systems'' in Ministry of Education, Culture, Sports,
Science and Technology (MEXT) (to D.M.) are also acknowledged. 

\appendix
\section*{[O I] 5577 as Abundance Indicator}

In this study, we had to invoke only one forbidden [O~{\sc i}] line 
of low excitation at 5577.34~$\rm\AA$ 
(2p$^{4}$~$^{1}$D$_{2}$--2p$^{4}$~$^{1}$S$_{0}$, $\chi_{\rm low}$ = 1.97~eV) 
for O-abundance determination, since this was the only available line 
in our spectrum data covering the 5000--6200~$\rm\AA$ region.\footnote{ 
Although we searched for the high-excitation O~{\sc i} 6155--58 lines 
as another possibility, they were too weak to be detected.} 
Having compared the resulting oxygen abundances ([O/H]$_{5577}$) with 
those of Takeda et al. (1998) derived from the O~{\sc i} 7771--5 
lines ([O/H]$_{7773}^{\rm NLTE}$)\footnote{We used 
$A_{\odot}^{\rm O,NLTE} = 8.82$ (Takeda \& Honda 2005) 
as the reference solar oxygen abundance for [O/H]$_{7773}^{\rm NLTE}$. 
Therefore, since the [O/H] values given in table 1
of Takeda et al. (1998) are the abundances relative to $\beta$ Gem
($A_{\beta{\rm Gem}}^{\rm O,NLTE} = 8.88$), a correction of +0.06 should be 
added in order to convert them to the abundances relative to the Sun.} for 
the 12 stars in common, we found a significantly large 
systematic difference (by $\sim$~0.3--0.4 dex; 0.37~dex on the average) 
between these two (the former is lower than the latter) as shown in figure 15a. 
Furthermore, when compared with Luck and Heiter's (2007) [O/H]$_{6300}$ 
results derived from the [O~{\sc i}] 6300.31~$\rm\AA$ line, a similar 
discrepancy was again recognized (figure 15b), which makes us suspect
that our [O/H]$_{5577}$ may be considerably underestimated.

The error in the $gf$ value (if any exists) is not relevant here, 
because our analysis is purely differential relative to the Sun. 
Also, it can not be due to the [O I] 5577 emission line of geo-atmospheric 
origin (which is surely observed in our spectrum), since its wavelength 
is generally different from that of the stellar line due to the Doppler 
shift (we anyhow gave up its measurement when an overlapping was confirmed 
by eye-inspection).

More strangely, when it comes to F--G--K dwarfs, [O/H]$_{5577}$ 
and [O/H]$_{7773}$ are consistent with each other as we can see in figures 
15c and d (though the uncertainties in the former are larger because of 
the difficulty in measuring weak lines), which means that the large 
discrepancy ([O/H]$_{5577} < $[O/H]$_{7773}$) occurs {\it only in giants}. 

As a possibility for explaining this confusing situation, we speculated 
that ``the [O~{\sc i}] 5577 line is significantly contaminated (even if 
superficially undetectable) by some blending component in solar-type dwarfs 
(including the Sun), whereas this blending effect becomes insignificant 
in the condition of low-gravity giants.'' If this is really the case, 
while the resulting overestimated solar oxygen abundance would cause 
an underestimation of [O/H]$_{5577}^{\rm giants}$, the 
[O/H]$_{5577}^{\rm dwarfs}$ would not be essentially affected because 
the error (acting on both the star and the Sun) is cancelled each other. 

Following this consideration, we searched Kurucz and Bell's (1995) list 
of spectral lines in the neighborhood of 5577.34~$\rm\AA$ and found 
that the Y~{\sc i} line at 5577.42~$\rm\AA$ can have an appreciable 
contribution. In order to examine whether its blending produces any 
quantitatively significant effect, 
we carried out spectrum synthesis analyses of the [O~{\sc i}]~5577 
region to find the best-fit O-abundance solutions of the Sun and HD~28305 
($\epsilon$ Tau; selected as a representative giant star) for the two cases: 
(1) both O and Y abundances are varied while including the Y~{\sc i} line, 
and (2) only O abundance is varied while neglecting the Y~{\sc i} line.
The resulting solutions of $\log\epsilon$(O) for cases (1)/(2) are 
8.99/9.05 (Sun) and 8.95/9.02 (HD~28305), and the appearance of the final 
fit between observed and theoretical spectra is depicted in figures 15e
(Sun) and f (HD~28305). We may conclude from these results that, 
although this Y~{\sc i} line shows some contribution to the absorption 
feature at $\sim 5577$~$\rm\AA$ (its inclusion surely improves the fitting),
its effect is insignificant in the quantitative sense (the extent of the 
abundance change is only $\ltsim 0.1$~dex) and thus can not be the cause
of the discrepancy amounting to $\sim$~0.3--0.4~dex.

Consequently, we could not find any reasonable solution to the problem of
why our [O/H] results derived from [O~{\sc i}] 5577 for late-G giants 
tend to be markedly lower than those based on O~{\sc i} 7771--5 or 
[O~{\sc i}]~6300. Further intensive studies toward clarifying the cause
of this disagreement (such as searching for some other blending candidate
on an updated line list, or investigating the line-formation 
mechanism\footnote{According to the conventional non-LTE calculation 
(e.g., Takeda et al. 1998) using ordinary plain-parallel atmospheric models, 
the formation of the [O~{\sc i}] 5577 line is almost perfectly described 
in LTE; i.e., no emission line is produced.}
in the presence of the extended circumstellar gas in order to search for a
possibility of filled-in emission which may lead to a weakening of absorption) 
would be required to settle this puzzling situation.  

In any case, modestly speaking, our results on oxygen abundances should be
viewed with caution since they may be systematically underestimated, 
though we would not conclude them to be totally erroneous  
as long as a possibility still exists (even if marginal) that 
the results from the other lines are overestimated by some unknown
reasons (e.g., blending effect in [O~{\sc i}] 6300 occurring only in giants, 
intensification of O~{\sc i} 7771--5 lines caused by chromospheric 
temperature rise).

\setcounter{figure}{14}
\begin{figure}
  \begin{center}
    \FigureFile(80mm,140mm){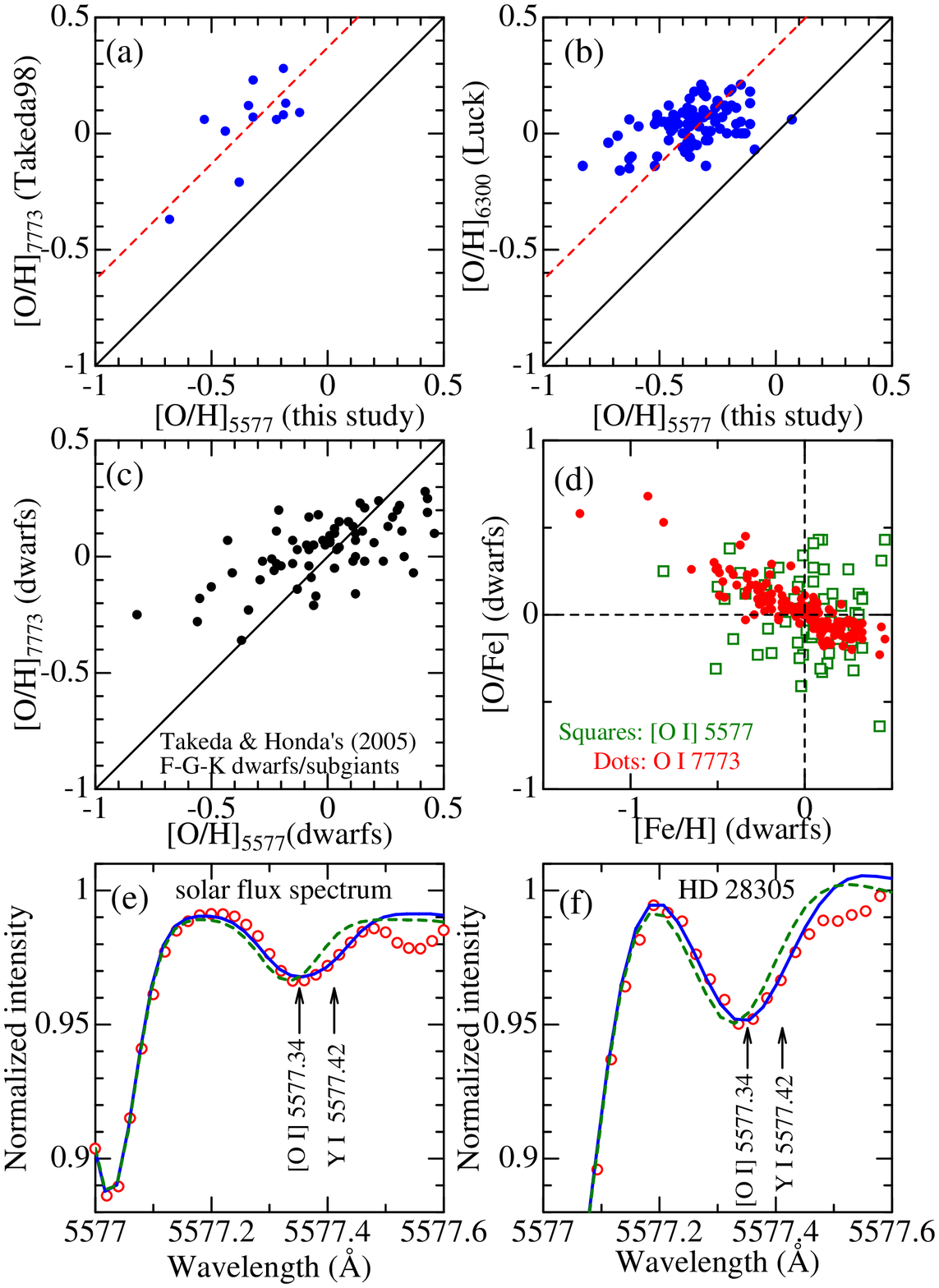}
  \end{center}
\caption{
(a) [O/H]$^{\rm NLTE}_{7773}$ (from O~{\sc i} 7771--5; Takeda et al. 1998) vs. 
[O/H]$_{5577}$ (from [O~{\sc i}] 5577; this study) correlation for 
12 stars in common. The relation 
$\langle$[O/H]$^{\rm NLTE}_{7773}\rangle - \langle$[O/H]$_{5577}\rangle = -0.37$
holds between these two averages, as shown by the dashed line.
(b) Comparison of [O/H]$_{6300}$ determined by Luck and Heiter (1997)
from the [O~{\sc i}] 6300 line with [O/H]$_{5577}$ derived in this study
based on the [O~{\sc i}] 5577 line, for 93 stars in common.
The dashed line indicates the relation [O/H]$_{6300}$ = [O/H]$_{5577}$ + 0.37
tentatively drawn in analogy with panel (a).
(c) Comparison of [O/H]$_{7773}^{\rm NLTE}$ and [O/H]$_{5577}$ for F--G--K 
dwarfs. The former is the non-LTE abundance derived from
O~{\sc i} 7771--5 triplet lines taken from table 2 of Takeda and Honda 
(2005), while the latter is the (LTE) abundance from the [O~{\sc i}] 5577 line
newly determined for this study based on Takeda et al.'s (2005a) spectra 
database (measurable for 70 objects out of 160 stars available).  
(d) [O/Fe] vs. [Fe/H] relation for F--G--K dwarfs.
Open squares represent [O/Fe]$_{5577}$ for 70 stars described above, 
while filled circles correspond to [O/Fe]$_{7773}^{\rm NLTE}$ of
160 stars (i.e., the same as figure 6c of Takeda \& Honda 2005).
(e) Spectrum fitting of the solar flux spectrum (Kurucz et al. 1984) 
in the 5577.0--5577.6~$\rm\AA$ region comprising [O~{\sc i}] 5577.34 and 
Y~{\sc i} 5577.42 lines. Open circles represent the observed spectrum,
while the best-fit theoretical spectra for two cases of different treatment
for the Y~{\sc i} line are shown by the solid line (Y~{\sc i} line included) 
and the dashed line (Y~{\sc i} line neglected). The strong feature at 
$\lambda \sim 5577$~$\rm\AA$ is due to Fe~{\sc i} 5577.03.
(f) Spectrum fitting of HD~28305 ($\epsilon$ Tau) in the 
5577.0--5577.6~$\rm\AA$ region. Otherwise, the same as in panel (e).
}
\end{figure}

\newpage

\onecolumn

\setcounter{figure}{10}
\begin{figure}
  \begin{center}
    \FigureFile(160mm,200mm){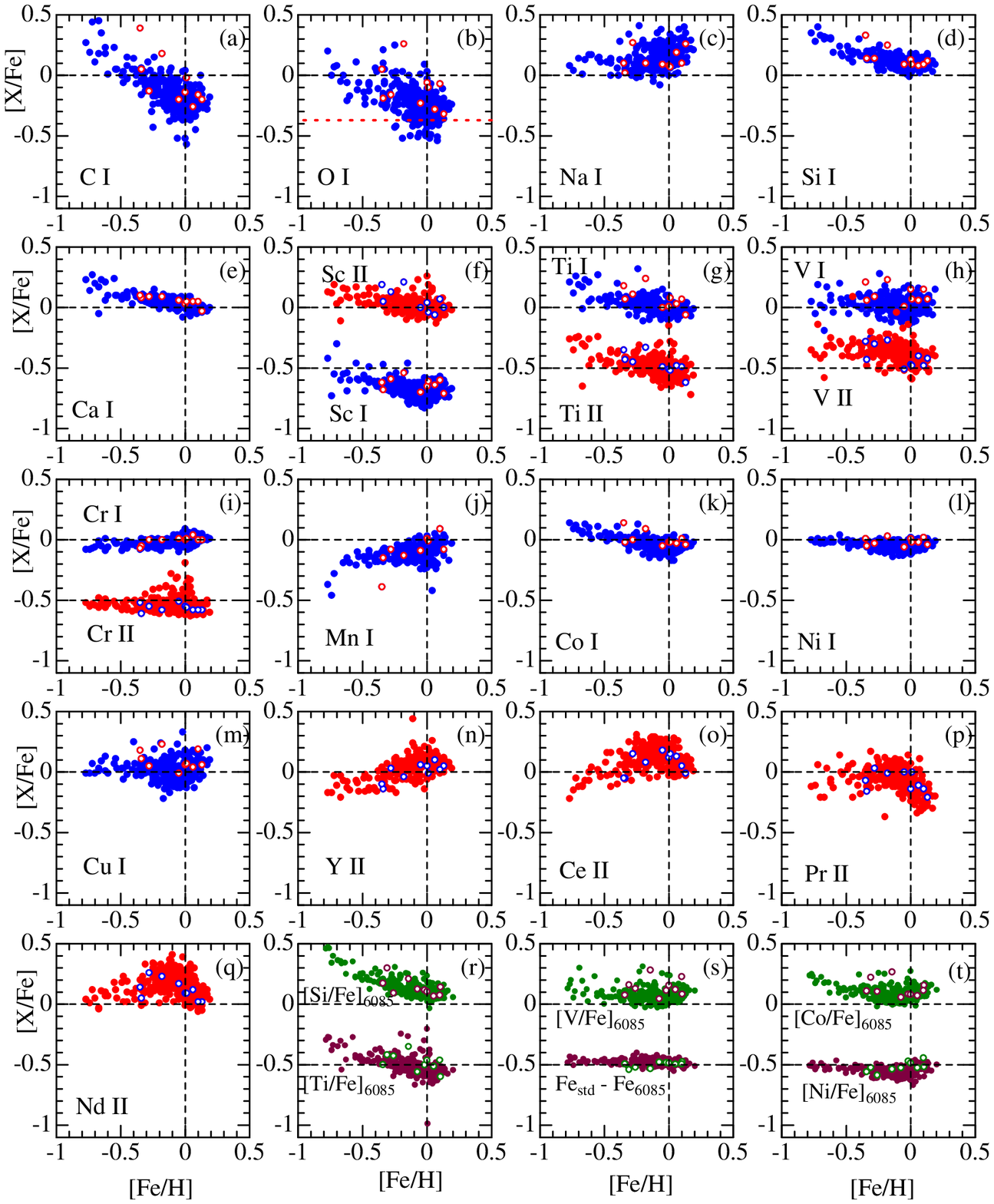}
  \end{center}
\caption{
In panels (a) through (q), [X/Fe] values ($\equiv$ [X/H] $-$ [Fe/H]
for each element X) derived from our abundance analyses using the 
measured equivalent widths are plotted against [Fe/H].
(a) [C/Fe], (b) [O/Fe] (see Appendix for the meaning of the horizontal
dotted line at [O/Fe] = $-0.37$, down to which the zero-point
might be lowered), (c) [Na/Fe], (d) [Si/Fe], (e) [Ca/Fe], 
(f) [Sc/Fe], (g) [Ti/Fe], (h) [V/Fe], (i) [Cr/Fe], (j) [Mn/Fe], 
(k) [Co/Fe], (l) [Ni/Fe], (m) [Cu/Fe], (n) [Y/Fe], (o) [Ce/Fe],
(p) [Pr/Fe], and (q) [Nd/Fe]. For Sc, Ti, V, and Cr, two kinds of 
results from lines of different ionization stages are separately 
shown in panels (f), (g), (h), and (i), respectively, where the lower 
results (vertically offset by $-0.5$~dex) may be comparatively less
reliable because of being based on smaller number of lines.
Meanwhile, the [X/Fe]$_{6085}$ values corresponding to the abundances 
($A_{6085}^{\rm Si}$, $A_{6085}^{\rm Ti}$, $A_{6085}^{\rm V}$, 
$A_{6085}^{\rm Fe}$, $A_{6085}^{\rm Co}$, and $A_{6085}^{\rm Ni}$) 
derived from the spectrum fitting in the 6080--6089~$\rm\AA$ region
are plotted against [Fe/H]$_{6085}$ in the last three panels: 
(r) [Si/Fe]$_{6085}$ and [Ti/Fe] (offset by $-0.5$~dex), 
(s) [V/Fe]$_{6085}$ and $A_{\rm std}^{\rm Fe} - A_{6085}^{\rm Fe}$
(offset by $-0.5$~dex), and (t) [Co/Fe]$_{6085}$ and [Ni/Fe] 
(offset by $-0.5$~dex). In all panels, 10 planet-host stars are
indicated by open symbols.
}
\end{figure}

\setcounter{table}{0}
\small
\renewcommand{\arraystretch}{0.8}
\setlength{\tabcolsep}{3pt}
\begin{longtable}{rlccccrrrrrrrcrcc}
\caption{Basic data and the parameter solutions of the program stars.}
\hline\hline
HD & Sp. & $V$ & $T_{\rm eff}$ & $\log g$ & $v_{\rm t}$ &
[Fe/H] & $\pi$ & $\sigma_{\pi} / \pi$ & $A_{V}$ & $M_{V}$ & B.C. &
$\log L$ & $M$ & $\log age$ & $\log g_{TLM}$ & Rem. \\
\hline
\endhead
\hline
\endfoot
\hline
\multicolumn{15}{l}{\hbox to 0pt{\parbox{180mm}{\footnotesize
Note. \\
The basic stellar data in columns 1--3 are self$-$explanatory, 
which were taken from the Hipparcos catalogue.
The values of $T_{\rm eff}$ (in K) , $\log g$ (in cm~s$^{-2}$), 
$v_{\rm t}$ (in km~s$^{-1}$), and [Fe/H] given in columns 4--7
are the finally established solutions based on our spectroscopic 
method using Fe~{\sc i} and Fe~{\sc ii} lines. Columns 8--15 gives 
the Hipparcos parallax ($\pi$; in unit of m.a.s.) along with the 
fractional error ($\sigma_{\pi}/\pi$) involved (ESA 1997), the 
estimated interstellar extinction ($A_{V}$), the absolute visual 
magnitude ($M_{V}$), the bolometric correction (B.C.), the stellar 
luminosity ($\log L/L_{\odot}$), the stellar mass ($M/M_{\odot}$), 
the stellar age ($\log age$, in yr) and the 
theoretical surface gravity ($\log g_{TLM}$, in cm~s$^{-2}$). 
See the text (sections 2 and 3) for more details. The planet-host 
stars are indicated by ``PHS'' in column 17, where ``PHS (BD)'' for
HD~107383 means that the companion is considered to be a brown dwarf.
}}}
\endlastfoot
\hline
    87&G5~III   &  5.55&  5072&  2.63&  1.35& $-$0.07&   8.8&  0.09&  0.07& +0.19& $-$0.24&  1.92&  2.74&  8.66&  2.73&     \\
   360&G8~III:  &  5.99&  4850&  2.62&  1.34& $-$0.08&   9.8&  0.09&  0.10& +0.84& $-$0.32&  1.69&  2.34&  8.86&  2.82&     \\
   448&G9~III   &  5.57&  4780&  2.51&  1.32& +0.03&  11.2&  0.06&  0.12& +0.69& $-$0.34&  1.76&  2.25&  8.99&  2.70&     \\
   587&K1~III   &  5.84&  4893&  3.08&  1.13& $-$0.09&  18.2&  0.05&  0.10& +2.04& $-$0.30&  1.20&  1.58&  9.36&  3.15&     \\
   645&K0~III   &  5.84&  4880&  3.03&  1.18& +0.07&  15.3&  0.05&  0.10& +1.67& $-$0.30&  1.35&  1.95&  9.08&  3.08&     \\
  1239&G8~III   &  5.74&  5114&  2.21&  1.63& $-$0.24&   5.1&  0.11&  0.49& $-$1.22& $-$0.23&  2.48&  3.75&  8.28&  2.32&     \\
  2114&G5~III   &  5.77&  5230&  2.57&  1.57& $-$0.03&   5.5&  0.19&  0.10& $-$0.63& $-$0.19&  2.23&  3.29&  8.45&  2.55&     \\
  2952&K0~III   &  5.93&  4844&  2.67&  1.32& +0.00&   8.7&  0.08&  0.18& +0.44& $-$0.32&  1.85&  2.54&  8.76&  2.69&     \\
  3421&G5~III   &  5.45&  5287&  1.88&  2.14& $-$0.20&   3.2&  0.24&  0.24& $-$2.27& $-$0.18&  2.88&  4.43&  8.13&  2.05&     \\
  3546&G5~III...&  4.34&  4882&  2.09&  1.44& $-$0.67&  19.3&  0.04&  0.08& +0.69& $-$0.31&  1.75&  2.00&  8.95&  2.70&     \\
  3817&G8~III   &  5.30&  5041&  2.52&  1.40& $-$0.12&   9.5&  0.09&  0.11& +0.07& $-$0.25&  1.97&  2.81&  8.62&  2.68&     \\
  3856&G9~III-IV&  5.83&  4766&  2.28&  1.35& $-$0.15&   6.5&  0.09&  0.42& $-$0.53& $-$0.35&  2.25&  3.09&  8.55&  2.34&     \\
  4188&K0~IIIvar&  4.77&  4844&  2.58&  1.32& $-$0.01&  15.5&  0.05&  0.10& +0.63& $-$0.32&  1.77&  2.54&  8.75&  2.76&     \\
  4398&G8/K0~III&  5.49&  4892&  2.56&  1.37& $-$0.18&   9.8&  0.07&  0.10& +0.34& $-$0.30&  1.88&  2.59&  8.72&  2.68&     \\
  4440&K0~IV    &  5.86&  4842&  2.91&  1.15& $-$0.10&  14.7&  0.09&  0.04& +1.65& $-$0.32&  1.37&  1.81&  9.19&  3.02&     \\
  4627&G8~III   &  5.92&  4599&  2.05&  1.40& $-$0.20&   4.9&  0.17&  0.12& $-$0.74& $-$0.43&  2.37&  3.06&  8.56&  2.16&     \\
  4732&K0~III   &  5.90&  4959&  3.16&  1.12& +0.01&  17.7&  0.06&  0.10& +2.04& $-$0.27&  1.19&  1.74&  9.24&  3.22&     \\
  5395&G8~III-IV&  4.62&  4774&  2.17&  1.40& $-$0.45&  15.8&  0.04&  0.18& +0.44& $-$0.35&  1.86&  1.95&  9.08&  2.54&     \\
  5608&K0      &  5.99&  4854&  3.03&  1.08& +0.06&  17.2&  0.05&  0.06& +2.11& $-$0.31&  1.18&  1.55&  9.40&  3.15&     \\
  5722&G7~III   &  5.62&  4893&  2.49&  1.39& $-$0.23&  10.3&  0.09&  0.10& +0.59& $-$0.30&  1.78&  2.26&  8.95&  2.72&     \\
  6186&K0~III   &  4.27&  4829&  2.30&  1.35& $-$0.31&  17.1&  0.05&  0.05& +0.39& $-$0.32&  1.88&  2.30&  8.92&  2.61&     \\
  7087&K0~III   &  4.66&  4908&  2.39&  1.53& $-$0.04&   7.4&  0.09&  0.14& $-$1.13& $-$0.29&  2.47&  3.83&  8.28&  2.27&     \\
  9057&K0~III   &  5.27&  4883&  2.49&  1.37& +0.04&  11.3&  0.07&  0.09& +0.44& $-$0.30&  1.85&  2.56&  8.78&  2.71&     \\
  9408&K0~III   &  4.68&  4746&  2.21&  1.40& $-$0.34&  16.0&  0.04&  0.18& +0.52& $-$0.36&  1.83&  2.04&  9.00&  2.57&     \\
  9774&G8~II-III&  5.28&  4980&  2.50&  1.60& +0.02&   7.3&  0.10&  0.09& $-$0.49& $-$0.27&  2.20&  3.25&  8.46&  2.49&     \\
 10348&K0~III   &  5.97&  4931&  2.55&  1.56& +0.01&   6.2&  0.13&  0.22& $-$0.28& $-$0.28&  2.12&  3.04&  8.54&  2.52&     \\
 10761&K0~III   &  4.26&  4952&  2.43&  1.43& $-$0.05&  12.6&  0.07&  0.07& $-$0.30& $-$0.28&  2.13&  3.04&  8.53&  2.52&     \\
 10975&K0~III   &  5.94&  4866&  2.47&  1.37& $-$0.17&  10.6&  0.07&  0.10& +0.96& $-$0.31&  1.64&  2.19&  8.94&  2.84&     \\
 11037&G9~III   &  5.91&  4862&  2.45&  1.33& $-$0.14&   9.9&  0.09&  0.08& +0.82& $-$0.31&  1.70&  2.30&  8.88&  2.80&     \\
 11949&K0~IV    &  5.70&  4845&  2.85&  1.17& $-$0.10&  13.2&  0.05&  0.11& +1.20& $-$0.32&  1.55&  2.17&  8.94&  2.92&     \\
 12139&K0~III-IV&  5.89&  4833&  2.53&  1.36& $-$0.09&   8.2&  0.10&  0.12& +0.33& $-$0.32&  1.89&  2.45&  8.87&  2.62&     \\
 12339&G8~III   &  5.22&  5011&  2.52&  1.51& $-$0.03&   7.7&  0.07&  0.08& $-$0.44& $-$0.26&  2.18&  3.19&  8.48&  2.51&     \\
 12583&K0~II/III&  5.87&  4969&  2.51&  1.45& +0.00&   9.9&  0.09&  0.10& +0.75& $-$0.27&  1.71&  2.48&  8.78&  2.86&     \\
 13468&G9~III:  &  5.94&  4893&  2.54&  1.34& $-$0.16&   9.2&  0.09&  0.09& +0.67& $-$0.30&  1.75&  2.31&  8.92&  2.76&     \\
 13692&K0~III   &  5.86&  4868&  2.55&  1.35& $-$0.12&   8.2&  0.10&  0.10& +0.32& $-$0.31&  1.90&  2.52&  8.80&  2.65&     \\
 13994&G7~III   &  5.99&  4974&  2.44&  1.83& $-$0.11&   4.6&  0.15&  0.52& $-$1.22& $-$0.27&  2.50&  3.84&  8.28&  2.27&     \\
 14129&G8~III   &  5.51&  4936&  2.61&  1.37& $-$0.01&   9.6&  0.10&  0.10& +0.32& $-$0.28&  1.88&  2.70&  8.68&  2.71&     \\
 14770&G8~III   &  5.19&  4977&  2.47&  1.47& +0.01&   8.7&  0.08&  0.17& $-$0.28& $-$0.27&  2.12&  3.03&  8.54&  2.54&     \\
 15779&G3~III:  &  5.36&  4846&  2.63&  1.26& +0.00&  12.3&  0.09&  0.07& +0.73& $-$0.32&  1.73&  2.49&  8.78&  2.79& \\
 15920&G8~III   &  5.17&  5061&  2.74&  1.33& $-$0.06&  12.7&  0.04&  0.29& +0.40& $-$0.24&  1.84&  2.63&  8.71&  2.79&     \\
 16400&G5~III:  &  5.65&  4785&  2.35&  1.33& $-$0.06&  10.3&  0.09&  0.08& +0.63& $-$0.34&  1.78&  2.43&  8.82&  2.71& \\
 16901&G0~Ib    &  5.43&  5624&  1.42&  3.17& +0.00&   4.7&  0.19&  0.37& $-$1.57& $-$0.11&  2.57&  4.03&  8.21&  2.43&     \\
 17656&G8~III   &  5.86&  5100&  2.67&  1.37& $-$0.06&   8.2&  0.10&  0.22& +0.21& $-$0.23&  1.91&  2.73&  8.66&  2.75&     \\
 17824&K0~III   &  4.76&  5051&  2.82&  1.19& $-$0.04&  17.9&  0.04&  0.10& +0.92& $-$0.24&  1.63&  2.37&  8.83&  2.95&     \\
 18474&G4p     &  5.47&  5013&  2.38&  1.42& $-$0.23&   5.8&  0.13&  0.30& $-$0.99& $-$0.26&  2.40&  3.59&  8.33&  2.35&     \\
 18953&K0~II-III&  5.32&  5029&  2.93&  1.23& +0.14&  12.7&  0.07&  0.10& +0.73& $-$0.25&  1.71&  2.53&  8.74&  2.89&     \\
 18970&K0~II-III&  4.77&  4791&  2.44&  1.30& $-$0.07&  15.9&  0.05&  0.19& +0.59& $-$0.34&  1.80&  2.44&  8.81&  2.70&     \\
 19476&K0~III   &  3.79&  4933&  2.82&  1.24& +0.14&  29.1&  0.02&  0.06& +1.05& $-$0.28&  1.59&  2.36&  8.83&  2.94&     \\
 19525&G9~III   &  6.28&  4801&  2.59&  1.38& $-$0.11&   7.0&  0.16&  0.08& +0.42& $-$0.33&  1.87&  2.37&  8.88&  2.63&     \\
 19845&G9~III   &  5.93&  4968&  2.86&  1.30& +0.14&  10.5&  0.06&  0.17& +0.86& $-$0.27&  1.67&  2.47&  8.77&  2.90&     \\
 20618&G8~IV    &  5.91&  5049&  3.08&  1.10& $-$0.22&  15.9&  0.07&  0.06& +1.85& $-$0.25&  1.26&  1.85&  9.12&  3.21&     \\
 20791&G8.5~III &  5.70&  4976&  2.63&  1.36& +0.07&  11.2&  0.07&  0.03& +0.92& $-$0.27&  1.64&  2.42&  8.81&  2.93&     \\
 20894&G8~III   &  5.50&  5119&  2.67&  1.44& $-$0.07&   7.8&  0.09&  0.16& $-$0.21& $-$0.22&  2.07&  2.97&  8.56&  2.63&     \\
 21755&G8~III   &  5.93&  5012&  2.45&  1.39& $-$0.13&   6.3&  0.15&  0.10& $-$0.17& $-$0.26&  2.07&  2.95&  8.56&  2.59&     \\
 22409&G7~III:  &  5.56&  5005&  2.67&  1.32& $-$0.25&   8.6&  0.09&  0.14& +0.09& $-$0.26&  1.97&  2.78&  8.62&  2.66&     \\
 22675&G5~III:  &  5.86&  4878&  2.50&  1.29& $-$0.06&   8.3&  0.09&  0.15& +0.32& $-$0.30&  1.89&  2.58&  8.76&  2.66&     \\
 22796&G6~III:  &  5.55&  4999&  2.72&  1.36& $-$0.10&   8.1&  0.10&  0.06& +0.04& $-$0.26&  1.99&  2.76&  8.67&  2.64&     \\
 23526&G9~III   &  5.91&  4837&  2.50&  1.30& $-$0.15&   9.7&  0.09&  0.04& +0.80& $-$0.32&  1.71&  2.27&  8.90&  2.78&     \\
 26409&G8~III   &  5.44&  5012&  2.67&  1.42& +0.03&   8.7&  0.09&  0.05& +0.08& $-$0.26&  1.97&  2.82&  8.62&  2.67&     \\
 27022&G5~III   &  5.26&  5314&  2.92&  1.29& $-$0.01&   9.8&  0.07&  0.25& $-$0.03& $-$0.17&  1.98&  2.84&  8.62&  2.76&     \\
 27348&G8~III   &  4.93&  5001&  2.75&  1.26& +0.05&  14.4&  0.06&  0.45& +0.27& $-$0.26&  1.90&  2.74&  8.66&  2.73&     \\
 27371&G8~III   &  3.65&  4923&  2.57&  1.34& +0.10&  21.2&  0.06&  0.06& +0.22& $-$0.29&  1.93&  2.80&  8.63&  2.68&     \\
 27697&G8~III   &  3.77&  4984&  2.64&  1.38& +0.12&  21.3&  0.04&  0.06& +0.35& $-$0.27&  1.87&  2.73&  8.66&  2.75&     \\
 27971&K1~III   &  5.29&  4886&  2.62&  1.31& +0.05&  13.4&  0.06&  0.48& +0.45& $-$0.30&  1.84&  2.56&  8.77&  2.71&     \\
 28100&G8~III   &  4.69&  5011&  2.54&  1.59& $-$0.08&   7.2&  0.11&  0.24& $-$1.27& $-$0.26&  2.51&  3.94&  8.23&  2.28&     \\
 28305&K0~III   &  3.53&  4883&  2.57&  1.46& +0.13&  21.0&  0.04&  0.06& +0.09& $-$0.30&  1.99&  2.84&  8.62&  2.62& PHS \\
 28307&G7~III   &  3.84&  4924&  2.63&  1.24& +0.10&  20.7&  0.04&  0.08& +0.34& $-$0.29&  1.88&  2.73&  8.66&  2.72&     \\
 29737&G6/G8~III&  5.56&  4858&  2.33&  1.37& $-$0.45&  10.3&  0.07&  0.21& +0.42& $-$0.31&  1.86&  2.23&  8.93&  2.63&     \\
 30557&G9~III   &  5.64&  4859&  2.57&  1.31& $-$0.02&  10.2&  0.07&  0.34& +0.33& $-$0.31&  1.89&  2.51&  8.84&  2.65&     \\
 30814&K0~III   &  5.03&  4842&  2.54&  1.34& $-$0.02&  13.4&  0.05&  0.18& +0.49& $-$0.32&  1.83&  2.49&  8.80&  2.70&     \\
 32008&G4~V     &  5.39&  5235&  3.21&  1.10& $-$0.25&  18.3&  0.06&  0.06& +1.64& $-$0.20&  1.32&  1.96&  9.04&  3.24&     \\
 33833&G7~III   &  5.90&  4963&  2.67&  1.33& $-$0.04&   7.3&  0.10&  0.13& +0.09& $-$0.27&  1.97&  2.78&  8.65&  2.65&     \\
 34538&G8~IV    &  5.48&  4809&  2.86&  1.08& $-$0.39&  20.7&  0.04&  0.05& +2.01& $-$0.33&  1.23&  1.25&  9.65&  2.99&     \\
 34559&G8~III   &  4.96&  4998&  2.74&  1.36& +0.00&  15.8&  0.05&  0.24& +0.72& $-$0.26&  1.72&  2.49&  8.77&  2.87&     \\
 35369&G8~III   &  4.13&  4852&  2.44&  1.33& $-$0.25&  18.7&  0.04&  0.06& +0.43& $-$0.31&  1.85&  2.33&  8.90&  2.65&     \\
 35410&K0~III   &  5.07&  4809&  2.58&  1.17& $-$0.33&  18.9&  0.04&  0.05& +1.41& $-$0.33&  1.47&  1.69&  9.24&  2.88&     \\
 36079&G5~II    &  2.81&  5209&  2.45&  1.62& $-$0.25&  20.5&  0.04&  0.05& $-$0.68& $-$0.20&  2.25&  3.29&  8.43&  2.52&     \\
 37160&G8~III-IV&  4.09&  4704&  2.49&  1.19& $-$0.65&  28.1&  0.03&  0.03& +1.30& $-$0.38&  1.53&  1.08&  9.87&  2.58&     \\
 38527&G8~III   &  5.78&  5046&  2.77&  1.19& $-$0.11&  10.9&  0.08&  0.07& +0.89& $-$0.25&  1.64&  2.36&  8.83&  2.93&     \\
 38656&G8~III   &  4.51&  4901&  2.50&  1.35& $-$0.17&  15.3&  0.05&  0.07& +0.37& $-$0.30&  1.87&  2.57&  8.74&  2.69&     \\
 39004&G7~III:  &  5.60&  5002&  2.67&  1.39& +0.04&   8.7&  0.11&  0.20& +0.09& $-$0.26&  1.97&  2.83&  8.62&  2.67&     \\
 39007&G8~III   &  5.79&  4994&  2.69&  1.16& +0.08&   9.8&  0.09&  0.08& +0.66& $-$0.26&  1.74&  2.55&  8.74&  2.85&     \\
 39019&G9~III:  &  5.54&  4964&  2.91&  1.30& +0.19&  10.4&  0.10&  0.07& +0.56& $-$0.27&  1.79&  2.65&  8.69&  2.81&     \\
 39364&G8~III/IV&  3.76&  4593&  2.30&  1.18& $-$0.72&  29.1&  0.02&  0.04& +1.04& $-$0.43&  1.66&  0.94& 10.03&  2.36&     \\
 41361&G7~III:  &  5.67&  4921&  2.12&  1.82& $-$0.08&   3.0&  0.29&  0.31& $-$2.28& $-$0.29&  2.93&  4.79&  8.05&  1.91&     \\
 41597&G8~III   &  5.35&  4494&  1.69&  1.46& $-$0.62&   9.3&  0.07&  0.05& +0.15& $-$0.49&  2.04&  1.34&  9.58&  2.09&     \\
 43023&G8~III   &  5.83&  5005&  2.71&  1.30& $-$0.10&  10.4&  0.07&  0.06& +0.85& $-$0.26&  1.66&  2.38&  8.82&  2.90&     \\
 43039&G8~IIIvar&  4.32&  4726&  2.27&  1.40& $-$0.32&  19.3&  0.04&  0.05& +0.70& $-$0.37&  1.77&  1.86&  9.12&  2.59&     \\
 45410&K0~IV    &  5.86&  4978&  3.16&  1.10& $-$0.13&  17.6&  0.04&  0.03& +2.05& $-$0.27&  1.19&  1.71&  9.24&  3.23& \\
 45415&G9~III   &  5.55&  4753&  2.39&  1.33& $-$0.12&  11.2&  0.07&  0.11& +0.68& $-$0.36&  1.77&  1.96&  9.12&  2.62&     \\
 46241&K0~V     &  5.88&  4919&  2.57&  1.37& $-$0.05&   6.3&  0.15&  0.19& $-$0.31& $-$0.29&  2.14&  3.04&  8.53&  2.50&     \\
 46480&G8~IV-V  &  5.94&  4866&  3.04&  1.05& $-$0.55&  18.8&  0.04&  0.03& +2.28& $-$0.31&  1.11&  1.08&  9.94&  3.06&     \\
 48432&K0~III   &  5.34&  4891&  2.83&  1.19& $-$0.11&  15.7&  0.05&  0.03& +1.29& $-$0.30&  1.51&  2.12&  8.97&  2.97&     \\
 50522&G5~III-IV&  4.35&  4850&  2.71&  0.96& +0.04&  19.1&  0.04&  0.03& +0.73& $-$0.31&  1.73&  2.50&  8.77&  2.80&     \\
 51000&G5~III   &  5.91&  5203&  2.94&  1.33& $-$0.06&   8.5&  0.11&  0.05& +0.50& $-$0.20&  1.78&  2.56&  8.73&  2.89&     \\
 51814&G8~III   &  5.96&  4846&  2.23&  1.57& $-$0.02&   3.5&  0.25&  0.16& $-$1.47& $-$0.32&  2.61&  4.17&  8.16&  2.14&     \\
 53329&G8~IV    &  5.55&  4888&  2.35&  1.35& $-$0.48&  10.7&  0.08&  0.01& +0.68& $-$0.30&  1.75&  1.99&  9.05&  2.70&     \\
 54131&G8~III   &  5.47&  4737&  2.37&  1.31& $-$0.18&  10.7&  0.08&  0.08& +0.53& $-$0.36&  1.83&  2.26&  8.91&  2.62&     \\
 54810&K0~III   &  4.91&  4703&  2.48&  1.23& $-$0.32&  15.4&  0.05&  0.04& +0.81& $-$0.38&  1.73&  2.09&  8.96&  2.68&     \\
 55730&G6~III   &  5.71&  4810&  2.47&  1.38& $-$0.17&   9.8&  0.09&  0.08& +0.60& $-$0.33&  1.79&  2.33&  8.86&  2.70&     \\
 57478&G8/K0~III&  5.59&  5032&  2.35&  1.43& $-$0.04&   5.9&  0.13&  0.12& $-$0.69& $-$0.25&  2.28&  3.42&  8.40&  2.46&     \\
 57727&G8~III   &  5.04&  5001&  2.89&  1.18& $-$0.12&  21.2&  0.04&  0.00& +1.67& $-$0.26&  1.34&  1.96&  9.07&  3.14&     \\
 58367&G8~III   &  4.99&  4911&  1.76&  2.04& $-$0.14&   3.3&  0.27&  0.20& $-$2.62& $-$0.29&  3.07&  4.78&  8.03&  1.77&     \\
 60986&K0~III   &  5.58&  5059&  2.78&  1.31& +0.03&  10.7&  0.09&  0.01& +0.71& $-$0.24&  1.71&  2.50&  8.77&  2.89&     \\
 61363&K0~III   &  5.58&  4762&  2.33&  1.38& $-$0.31&  10.0&  0.08&  0.05& +0.53& $-$0.35&  1.83&  2.12&  8.96&  2.60&     \\
 62345&G8~III   &  3.57&  4979&  2.58&  1.39& $-$0.06&  22.7&  0.04&  0.00& +0.35& $-$0.27&  1.87&  2.65&  8.70&  2.74&     \\
 62509&K0~IIIvar&  1.16&  4904&  2.84&  1.26& +0.06&  96.7&  0.01&  0.00& +1.09& $-$0.29&  1.58&  2.31&  8.86&  2.94& PHS \\
 64152&K0~III   &  5.62&  5017&  2.89&  1.18& +0.07&  11.9&  0.06&  0.06& +0.94& $-$0.25&  1.62&  2.41&  8.81&  2.95&     \\
 65228&F7/F8~II &  4.20&  5932&  1.96&  3.30& +0.01&   6.5&  0.11&  0.09& $-$1.83& $-$0.07&  2.66&  4.18&  8.16&  2.45&     \\
 65345&K0~III   &  5.30&  4983&  2.73&  1.27& $-$0.05&  12.3&  0.08&  0.00& +0.75& $-$0.27&  1.71&  2.45&  8.79&  2.86&     \\
 65714&G8~III:  &  5.87&  4923&  2.45&  1.53& +0.08&   2.9&  0.31&  0.44& $-$2.26& $-$0.29&  2.92&  4.91&  7.97&  1.93&     \\
 67447&G8~II    &  5.34&  4974&  2.12&  2.12& $-$0.06&   3.1&  0.21&  0.04& $-$2.26& $-$0.27&  2.91&  4.85&  8.03&  1.95&     \\
 68077&G9~III   &  5.88&  4881&  2.48&  1.48& $-$0.01&   6.6&  0.11&  0.07& $-$0.10& $-$0.30&  2.06&  2.89&  8.61&  2.55&     \\
 68290&K0~III   &  4.72&  5028&  2.92&  1.21& +0.06&  17.6&  0.04&  0.04& +0.91& $-$0.25&  1.64&  2.41&  8.81&  2.94&     \\
 68312&G8~III   &  5.36&  5037&  2.70&  1.30& $-$0.12&  10.3&  0.08&  0.03& +0.40& $-$0.25&  1.84&  2.61&  8.71&  2.78&     \\
 68375&G8~III   &  5.55&  5041&  2.77&  1.29& $-$0.09&  11.2&  0.05&  0.00& +0.79& $-$0.25&  1.68&  2.42&  8.80&  2.90&     \\
 71088&G8~III   &  5.89&  4944&  2.75&  1.33& $-$0.07&  10.1&  0.06&  0.00& +0.92& $-$0.28&  1.64&  2.33&  8.85&  2.89&     \\
 71115&G8~II    &  5.13&  5062&  2.53&  1.49& $-$0.07&   9.1&  0.10&  0.00& $-$0.08& $-$0.24&  2.03&  2.90&  8.59&  2.64&     \\
 71369&G4~II-III&  3.35&  5242&  2.64&  1.51& $-$0.09&  17.8&  0.04&  0.00& $-$0.40& $-$0.19&  2.14&  3.09&  8.51&  2.62&     \\
 73017&G8~IV    &  5.66&  4735&  2.44&  1.20& $-$0.54&  13.6&  0.05&  0.06& +1.26& $-$0.37&  1.54&  1.31&  9.63&  2.67&     \\
 73593&G0~IV    &  5.35&  4755&  2.62&  1.16& $-$0.23&  18.1&  0.04&  0.06& +1.58& $-$0.35&  1.41&  1.48&  9.45&  2.86&     \\
 74395&G2~Ib    &  4.63&  5257&  1.68&  2.47& $-$0.07&   5.4&  0.18&  0.17& $-$1.88& $-$0.19&  2.73&  4.44&  8.08&  2.20&     \\
 74739&G8~Iab:  &  4.03&  4905&  2.25&  1.80& $-$0.06&  10.9&  0.12&  0.00& $-$0.77& $-$0.29&  2.33&  3.43&  8.42&  2.36&     \\
 74918&G8~III   &  4.32&  5063&  2.70&  1.34& $-$0.14&  14.4&  0.11&  0.12& $-$0.01& $-$0.24&  2.00&  2.86&  8.60&  2.66&     \\
 75506&K0~III   &  5.15&  4811&  2.35&  1.38& $-$0.35&  11.9&  0.06&  0.07& +0.46& $-$0.33&  1.85&  2.05&  9.06&  2.58&     \\
 76219&G8~II-III&  5.23&  4904&  2.13&  1.75& $-$0.15&   5.7&  0.15&  0.06& $-$1.06& $-$0.29&  2.44&  3.59&  8.37&  2.27&     \\
 76294&G8~III-IV&  3.11&  4844&  2.30&  1.41& $-$0.11&  21.6&  0.05&  0.03& $-$0.24& $-$0.32&  2.12&  2.93&  8.60&  2.48&     \\
 76813&G9~III   &  5.23&  5043&  2.64&  1.33& $-$0.06&  10.2&  0.07&  0.00& +0.27& $-$0.25&  1.89&  2.70&  8.67&  2.74&     \\
 77912&G8~Ib-II &  4.56&  4899&  1.75&  2.13& $-$0.14&   4.8&  0.16&  0.07& $-$2.10& $-$0.30&  2.86&  4.60&  8.08&  1.96&     \\
 78235&G8~III   &  5.42&  5123&  3.00&  1.19& $-$0.03&  12.6&  0.06&  0.00& +0.91& $-$0.22&  1.63&  2.38&  8.83&  2.98&     \\
 78668&G6~III   &  5.76&  5020&  2.74&  1.28& $-$0.07&   7.1&  0.13&  0.16& $-$0.15& $-$0.25&  2.06&  2.94&  8.57&  2.60&     \\
 79181&G8~III   &  5.72&  4842&  2.47&  1.35& $-$0.29&  10.8&  0.07&  0.11& +0.79& $-$0.32&  1.71&  2.10&  8.97&  2.74&     \\
 79452&G6~III   &  5.98&  4990&  2.27&  1.47& $-$0.74&   7.2&  0.13&  0.07& +0.19& $-$0.27&  1.93&  2.04&  9.01&  2.56&     \\
 80499&G8~III   &  4.77&  5033&  2.44&  1.46& $-$0.09&  10.2&  0.08&  0.12& $-$0.31& $-$0.25&  2.12&  3.05&  8.53&  2.56&     \\
 81688&K0~III-IV&  5.40&  4771&  2.26&  1.36& $-$0.34&  11.3&  0.07&  0.10& +0.57& $-$0.35&  1.81&  2.07&  8.98&  2.61& PHS \\
 82087&G8~III:  &  5.87&  4867&  2.59&  1.33& +0.02&   6.3&  0.14&  0.14& $-$0.26& $-$0.31&  2.13&  3.05&  8.53&  2.50&     \\
 82210&G4~III-IV&  4.54&  5299&  3.49&  1.13& $-$0.21&  30.9&  0.02&  0.00& +1.99& $-$0.18&  1.18&  1.81&  9.15&  3.37&     \\
 82734&K0~III   &  5.02&  4959&  2.62&  1.61& +0.17&   9.8&  0.07&  0.12& $-$0.15& $-$0.27&  2.07&  3.03&  8.52&  2.58&     \\
 82741&K0~III   &  4.81&  4801&  2.42&  1.31& $-$0.22&  14.2&  0.06&  0.08& +0.50& $-$0.34&  1.84&  2.17&  9.00&  2.62&     \\
 83506&K0~III   &  5.15&  4860&  2.36&  1.70& +0.07&   7.4&  0.08&  0.00& $-$0.51& $-$0.31&  2.23&  3.30&  8.44&  2.43&     \\
 83805&G8~III   &  5.61&  4997&  2.64&  1.33& +0.01&   9.6&  0.08&  0.11& +0.41& $-$0.26&  1.84&  2.65&  8.70&  2.77&     \\
 84441&G0~II    &  2.97&  5385&  2.18&  1.88& $-$0.09&  13.0&  0.07&  0.03& $-$1.49& $-$0.16&  2.56&  4.01&  8.21&  2.36&     \\
 85444&G6/G8~III&  4.11&  5045&  2.56&  1.35& $-$0.01&  11.9&  0.07&  0.10& $-$0.61& $-$0.25&  2.24&  3.34&  8.43&  2.48&     \\
 91190&K0~III   &  4.86&  4962&  2.59&  1.33& $-$0.03&  12.7&  0.04&  0.00& +0.38& $-$0.27&  1.86&  2.66&  8.70&  2.74&     \\
 91612&G8~II-III&  5.07&  4920&  2.55&  1.31& $-$0.20&  10.2&  0.08&  0.02& +0.10& $-$0.29&  1.98&  2.60&  8.78&  2.60&     \\
 92125&G0~II    &  4.68&  5468&  2.22&  2.07& +0.03&   6.9&  0.12&  0.05& $-$1.18& $-$0.14&  2.43&  3.72&  8.30&  2.48&     \\
 93291&G4~III:  &  5.49&  5039&  2.74&  1.28& $-$0.10&  11.3&  0.08&  0.02& +0.74& $-$0.25&  1.70&  2.43&  8.79&  2.88&     \\
 94402&G8~III   &  5.45&  4984&  2.64&  1.36& +0.03&  10.4&  0.08&  0.02& +0.53& $-$0.27&  1.79&  2.59&  8.73&  2.80&     \\
 94497&G7~III:  &  5.73&  4804&  2.69&  1.24& $-$0.14&  10.7&  0.07&  0.05& +0.82& $-$0.33&  1.71&  1.96&  9.12&  2.70&     \\
 95808&G7~III...&  5.51&  4935&  2.62&  1.29& $-$0.09&  10.2&  0.09&  0.11& +0.45& $-$0.28&  1.84&  2.58&  8.74&  2.74&     \\
 98839&G8~II    &  4.99&  4936&  2.30&  1.78& $-$0.05&   6.6&  0.10&  0.08& $-$0.98& $-$0.28&  2.41&  3.69&  8.33&  2.33&     \\
 99055&G8~IIICN.&  5.39&  5060&  2.65&  1.38& $-$0.05&   8.9&  0.09&  0.02& +0.12& $-$0.24&  1.95&  2.78&  8.64&  2.70&     \\
 99283&K0~III   &  5.73&  4883&  2.59&  1.37& $-$0.17&   9.4&  0.08&  0.00& +0.59& $-$0.30&  1.79&  2.29&  8.93&  2.72&     \\
 99648&G8~II-III&  4.95&  5002&  2.40&  1.63& $-$0.01&   5.2&  0.16&  0.02& $-$1.47& $-$0.26&  2.59&  4.21&  8.15&  2.22&     \\
100615&K0~III   &  5.63&  4827&  2.60&  1.34& $-$0.12&   7.9&  0.08&  0.00& +0.13& $-$0.32&  1.98&  2.57&  8.79&  2.56&     \\
100696&K0~III   &  5.19&  4833&  2.32&  1.35& $-$0.33&  13.5&  0.04&  0.00& +0.84& $-$0.32&  1.69&  2.01&  9.01&  2.74&     \\
100920&G9~III   &  4.30&  4835&  2.47&  1.32& $-$0.19&  18.3&  0.05&  0.02& +0.59& $-$0.32&  1.79&  2.23&  8.97&  2.69&     \\
101484&K1~III   &  5.26&  4893&  2.70&  1.24& +0.03&  14.0&  0.06&  0.04& +0.96& $-$0.30&  1.64&  2.38&  8.83&  2.89&     \\
102070&G8~III   &  4.71&  4992&  2.60&  1.51& +0.03&   9.3&  0.09&  0.21& $-$0.66& $-$0.26&  2.27&  3.42&  8.40&  2.45&     \\
103462&G8~III   &  5.26&  4903&  2.26&  1.39& $-$0.60&  11.1&  0.06&  0.19& +0.29& $-$0.30&  1.90&  2.14&  8.90&  2.58&     \\
103484&K0~III:  &  5.58&  5008&  3.18&  1.13& $-$0.01&  19.4&  0.04&  0.03& +1.99& $-$0.26&  1.21&  1.83&  9.18&  3.25&     \\
104979&G8~III   &  4.12&  4871&  2.48&  1.37& $-$0.45&  19.1&  0.04&  0.00& +0.52& $-$0.31&  1.82&  2.12&  9.00&  2.65&     \\
104985&G9~III   &  5.78&  4679&  2.47&  1.40& $-$0.35&   9.8&  0.05&  0.00& +0.74& $-$0.39&  1.76&  2.12&  8.94&  2.64& PHS \\
106057&K0~II-III&  5.60&  4956&  2.64&  1.35& $-$0.10&   6.7&  0.11&  0.06& $-$0.32& $-$0.28&  2.14&  3.06&  8.52&  2.52&     \\
106714&K0~III   &  4.93&  4933&  2.57&  1.37& $-$0.18&  13.1&  0.07&  0.05& +0.47& $-$0.28&  1.82&  2.50&  8.79&  2.74&     \\
107383&G8~III   &  4.72&  4841&  2.51&  1.38& $-$0.28&   9.0&  0.10&  0.05& $-$0.55& $-$0.32&  2.25&  3.14&  8.49&  2.38& PHS (BD) \\
107950&G7~III   &  4.76&  5171&  2.60&  1.63& +0.01&   8.3&  0.07&  0.04& $-$0.68& $-$0.21&  2.26&  3.36&  8.42&  2.52&     \\
108225&G8~III-IV&  5.01&  4969&  2.71&  1.27& +0.04&  14.3&  0.04&  0.04& +0.75& $-$0.27&  1.71&  2.50&  8.77&  2.87&     \\
109272&G8~III/IV&  5.58&  5104&  3.22&  1.13& $-$0.26&  20.6&  0.04&  0.13& +2.02& $-$0.23&  1.19&  1.79&  9.16&  3.29&     \\
109317&K0~IIICN.&  5.42&  4866&  2.61&  1.38& $-$0.05&  12.3&  0.06&  0.07& +0.79& $-$0.31&  1.71&  2.41&  8.82&  2.81&     \\
109379&G5~II    &  2.65&  5145&  2.56&  1.62& $-$0.01&  23.3&  0.03&  0.10& $-$0.61& $-$0.22&  2.23&  3.31&  8.44&  2.53&     \\
110646&G8~IIIp  &  5.91&  5067&  3.05&  1.21& $-$0.45&  14.3&  0.05&  0.00& +1.68& $-$0.24&  1.33&  1.81&  9.12&  3.14&     \\
111028&K1~III-IV&  5.65&  4881&  3.27&  1.03& $-$0.05&  22.4&  0.04&  0.00& +2.40& $-$0.30&  1.06&  1.41&  9.53&  3.24&     \\
113095&K0~III   &  5.97&  4961&  2.68&  1.37& $-$0.07&   8.1&  0.10&  0.07& +0.45& $-$0.27&  1.83&  2.59&  8.73&  2.76&     \\
113226&G8~IIIvar&  2.85&  5044&  2.63&  1.41& +0.07&  31.9&  0.03&  0.02& +0.35& $-$0.25&  1.86&  2.70&  8.68&  2.78&     \\
114256&K0~III   &  5.79&  4858&  2.68&  1.34& +0.04&   9.3&  0.08&  0.06& +0.57& $-$0.31&  1.80&  2.51&  8.77&  2.74&     \\
114946&G8~III/IV&  5.31&  5066&  3.32&  1.08& $-$0.33&  25.9&  0.03&  0.09& +2.29& $-$0.24&  1.08&  1.62&  9.28&  3.34&     \\
115202&K1~III   &  5.21&  4826&  3.11&  1.07& $-$0.02&  25.7&  0.03&  0.09& +2.17& $-$0.32&  1.16&  1.45&  9.50&  3.12&     \\
115659&G8~III   &  2.99&  5019&  2.47&  1.47& $-$0.06&  24.7&  0.03&  0.10& $-$0.15& $-$0.25&  2.06&  2.94&  8.57&  2.60&     \\
116292&K0~III   &  5.36&  4884&  2.49&  1.30& $-$0.09&  10.2&  0.07&  0.20& +0.20& $-$0.30&  1.94&  2.64&  8.72&  2.63&     \\
116957&K0~III:  &  5.88&  4898&  2.63&  1.33& $-$0.10&   9.1&  0.07&  0.02& +0.66& $-$0.30&  1.75&  2.39&  8.87&  2.78&     \\
117566&G2.5~IIIb&  5.74&  5496&  3.34&  1.37& +0.05&  11.2&  0.04&  0.00& +0.98& $-$0.13&  1.56&  2.29&  8.88&  3.15&     \\
117818&K0~III   &  5.21&  4811&  2.31&  1.34& $-$0.34&  12.4&  0.06&  0.17& +0.50& $-$0.33&  1.83&  2.05&  9.06&  2.60&     \\
118219&G6~III   &  5.70&  4831&  2.34&  1.33& $-$0.34&   8.8&  0.09&  0.17& +0.25& $-$0.32&  1.93&  2.51&  8.74&  2.60&     \\
119126&G9~III   &  5.63&  4796&  2.33&  1.34& $-$0.12&  10.1&  0.08&  0.07& +0.59& $-$0.34&  1.80&  2.38&  8.85&  2.69&     \\
119605&G1~IV/V  &  5.55&  5456&  1.96&  1.95& $-$0.31&   4.2&  0.17&  0.36& $-$1.70& $-$0.15&  2.64&  4.04&  8.19&  2.31&     \\
120048&G9~III   &  5.92&  5014&  2.79&  1.22& +0.11&   8.1&  0.08&  0.10& +0.36& $-$0.26&  1.86&  2.71&  8.66&  2.77&     \\
120084&G7~III:  &  5.91&  4892&  2.71&  1.31& +0.09&  10.2&  0.05&  0.00& +0.96& $-$0.30&  1.64&  2.39&  8.82&  2.89&     \\
120420&K0~III   &  5.61&  4791&  2.63&  1.26& $-$0.20&  10.5&  0.06&  0.10& +0.61& $-$0.34&  1.79&  2.25&  8.90&  2.68&     \\
120787&G3~V     &  5.97&  4843&  2.31&  1.34& $-$0.38&   8.3&  0.07&  0.00& +0.55& $-$0.32&  1.81&  2.02&  9.08&  2.63&     \\
125454&G9~III   &  5.14&  4848&  2.56&  1.39& $-$0.10&  11.9&  0.08&  0.10& +0.42& $-$0.32&  1.86&  2.47&  8.82&  2.67&     \\
126218&K0~III   &  5.34&  5025&  2.50&  1.58& +0.12&   8.2&  0.11&  0.26& $-$0.36& $-$0.25&  2.15&  3.15&  8.48&  2.55&     \\
127243&G3~IV    &  5.58&  4893&  2.21&  1.48& $-$0.77&  10.6&  0.06&  0.01& +0.69& $-$0.30&  1.74&  1.92&  9.08&  2.69&     \\
129312&G8~IIIvar&  4.86&  4993&  2.53&  1.62& +0.01&   5.7&  0.14&  0.05& $-$1.43& $-$0.26&  2.58&  4.20&  8.15&  2.23&     \\
129336&G8~III   &  5.55&  4901&  2.54&  1.33& $-$0.25&   8.5&  0.10&  0.05& +0.14& $-$0.30&  1.96&  2.68&  8.67&  2.62&     \\
129944&K0~III   &  5.80&  4892&  2.50&  1.32& $-$0.26&   8.9&  0.11&  0.25& +0.30& $-$0.30&  1.90&  2.59&  8.70&  2.66&     \\
129972&K0~III   &  4.60&  4976&  2.69&  1.43& $-$0.01&  14.5&  0.05&  0.05& +0.35& $-$0.27&  1.87&  2.68&  8.69&  2.74&     \\
130952&G8...   &  4.93&  4750&  2.34&  1.35& $-$0.40&  15.1&  0.07&  0.08& +0.74& $-$0.36&  1.75&  1.85&  9.10&  2.62&     \\
131530&G7~III   &  5.78&  4962&  2.72&  1.33& +0.00&   8.9&  0.11&  0.26& +0.28& $-$0.27&  1.90&  2.72&  8.67&  2.71&     \\
132146&G5~III:  &  5.72&  5012&  2.29&  1.60& $-$0.06&   5.3&  0.15&  0.05& $-$0.72& $-$0.26&  2.29&  3.45&  8.39&  2.44&     \\
133002&F9~V     &  5.63&  5532&  3.56&  1.11& $-$0.34&  23.1&  0.02&  0.00& +2.45& $-$0.14&  0.98&  1.49&  9.37&  3.56&     \\
133208&G8~III   &  3.49&  5001&  2.35&  1.61& $-$0.07&  14.9&  0.04&  0.06& $-$0.70& $-$0.26&  2.29&  3.42&  8.40&  2.44&     \\
133392&G8~III:  &  5.52&  4903&  2.69&  1.32& +0.09&  11.8&  0.05&  0.10& +0.79& $-$0.29&  1.70&  2.49&  8.77&  2.85&     \\
134190&G8~III   &  5.24&  4841&  2.28&  1.40& $-$0.41&  12.5&  0.04&  0.06& +0.67& $-$0.32&  1.76&  2.03&  8.99&  2.68&     \\
136512&K0~III   &  5.51&  4749&  2.34&  1.39& $-$0.29&  11.9&  0.06&  0.05& +0.84& $-$0.36&  1.71&  2.13&  8.93&  2.72& \\
136956&G8~III   &  5.72&  5031&  2.61&  1.54& +0.08&   5.4&  0.14&  0.44& $-$1.05& $-$0.25&  2.42&  3.78&  8.27&  2.36&     \\
138716&K1~IV    &  4.61&  4830&  3.14&  1.05& +0.00&  34.5&  0.02&  0.07& +2.23& $-$0.32&  1.14&  1.44&  9.51&  3.15&     \\
138852&K0~III-IV&  5.74&  4900&  2.55&  1.36& $-$0.22&  10.2&  0.05&  0.06& +0.73& $-$0.30&  1.73&  2.21&  8.98&  2.77&     \\
138905&K0~III   &  3.91&  4822&  2.56&  1.27& $-$0.30&  21.4&  0.04&  0.11& +0.45& $-$0.33&  1.85&  2.15&  9.01&  2.61& \\
139641&G8~III-IV&  5.25&  4907&  2.75&  1.16& $-$0.53&  20.0&  0.03&  0.09& +1.67& $-$0.30&  1.35&  1.43&  9.50&  2.96&     \\
141680&G8~III   &  5.21&  4770&  2.32&  1.34& $-$0.24&  12.4&  0.06&  0.19& +0.49& $-$0.35&  1.85&  2.17&  8.95&  2.60&     \\
142091&K0~III-IV&  4.79&  4877&  3.21&  1.04& +0.10&  32.1&  0.02&  0.03& +2.29& $-$0.30&  1.11&  1.51&  9.43&  3.22& PHS \\
142198&K0~III   &  4.13&  4760&  2.35&  1.39& $-$0.27&  20.0&  0.04&  0.25& +0.39& $-$0.35&  1.88&  2.13&  9.02&  2.55&     \\
142531&G8~III:  &  5.81&  4961&  2.78&  1.37& +0.05&   9.1&  0.06&  0.13& +0.47& $-$0.27&  1.82&  2.64&  8.70&  2.77&     \\
143553&K0~III:  &  5.82&  4805&  2.85&  1.17& $-$0.23&  13.6&  0.06&  0.18& +1.31& $-$0.33&  1.51&  1.75&  9.20&  2.85&     \\
144608&G6/G8~III&  4.31&  5266&  2.54&  1.60& $-$0.09&  12.3&  0.07&  0.38& $-$0.62& $-$0.19&  2.22&  3.27&  8.45&  2.57&     \\
145001&G8~III   &  5.00&  5119&  2.90&  1.57& +0.04&   8.4&  0.15&  0.07& $-$0.45& $-$0.22&  2.17&  3.17&  8.49&  2.56&     \\
146791&G8~III   &  3.23&  4931&  2.69&  1.34& $-$0.07&  30.3&  0.03&  0.09& +0.55& $-$0.28&  1.79&  2.52&  8.78&  2.77&     \\
147677&K0~III   &  4.86&  4978&  2.90&  1.28& +0.10&  17.8&  0.04&  0.07& +1.04& $-$0.27&  1.59&  2.36&  8.83&  2.96&     \\
147700&K0~III   &  4.48&  4843&  2.48&  1.31& $-$0.11&  18.3&  0.05&  0.27& +0.52& $-$0.32&  1.82&  2.35&  8.89&  2.69&     \\
148387&G8~III   &  2.73&  5055&  2.82&  1.34& $-$0.04&  37.2&  0.01&  0.03& +0.55& $-$0.24&  1.78&  2.55&  8.74&  2.84&     \\
148604&G5~III/IV&  5.66&  5120&  2.90&  0.98& $-$0.16&  12.2&  0.08&  0.44& +0.65& $-$0.23&  1.73&  2.48&  8.76&  2.89&     \\
148786&G8/K0~III&  4.29&  5110&  2.69&  1.52& +0.17&  15.5&  0.05&  0.36& $-$0.11& $-$0.23&  2.03&  2.96&  8.55&  2.66&     \\
150030&G8~II    &  5.83&  4850&  2.10&  1.81& $-$0.09&   3.7&  0.14&  0.10& $-$1.42& $-$0.32&  2.59&  4.02&  8.22&  2.14&     \\
150997&G8~III-IV&  3.48&  5045&  2.79&  1.26& $-$0.15&  29.1&  0.02&  0.05& +0.75& $-$0.25&  1.70&  2.41&  8.80&  2.89&     \\
152815&G8~III   &  5.39&  4859&  2.43&  1.35& $-$0.21&  12.8&  0.06&  0.07& +0.86& $-$0.31&  1.68&  2.19&  8.93&  2.80&     \\
154084&G7~III:  &  5.76&  4862&  2.62&  1.41& $-$0.16&   8.8&  0.07&  0.07& +0.42& $-$0.31&  1.86&  2.39&  8.89&  2.66&     \\
154779&K0~III   &  5.98&  5064&  2.75&  1.44& +0.12&   8.1&  0.11&  0.33& +0.20& $-$0.24&  1.92&  2.79&  8.63&  2.74&     \\
156874&K0~III   &  5.68&  4982&  2.85&  1.32& +0.00&  10.2&  0.06&  0.10& +0.63& $-$0.27&  1.75&  2.53&  8.76&  2.83&     \\
156891&G7~III:  &  5.97&  4981&  2.95&  1.30& +0.13&  10.2&  0.05&  0.10& +0.91& $-$0.27&  1.64&  2.44&  8.79&  2.93&     \\
157527&K0~III   &  5.82&  5090&  2.96&  1.30& +0.07&  10.8&  0.09&  0.25& +0.74& $-$0.23&  1.70&  2.49&  8.77&  2.92&     \\
158974&G8~III   &  5.63&  4901&  2.32&  1.43& $-$0.07&   8.7&  0.06&  0.13& +0.19& $-$0.30&  1.94&  2.74&  8.66&  2.65&     \\
159181&G2~II    &  2.79&  5153&  1.50&  2.69& $-$0.15&   9.0&  0.05&  0.10& $-$2.53& $-$0.22&  3.00&  4.65&  8.09&  1.91&     \\
159353&K0~III:  &  5.68&  4919&  2.76&  1.32& +0.00&  10.2&  0.08&  0.37& +0.35& $-$0.29&  1.88&  2.69&  8.69&  2.71&     \\
160781&G7~III   &  5.97&  4593&  2.10&  1.62& $-$0.02&   2.6&  0.30&  0.69& $-$2.62& $-$0.43&  3.12&  4.99&  8.00&  1.62&     \\
161178&G9~III   &  5.87&  4766&  2.33&  1.32& $-$0.20&  10.2&  0.05&  0.04& +0.87& $-$0.35&  1.69&  2.14&  8.94&  2.74&     \\
162076&G5~IV    &  5.69&  5018&  2.98&  1.24& +0.04&  13.0&  0.05&  0.09& +1.18& $-$0.25&  1.53&  2.27&  8.89&  3.02&     \\
163532&G9~III   &  5.44&  4689&  2.17&  1.44& $-$0.06&   7.7&  0.09&  0.53& $-$0.67& $-$0.39&  2.32&  3.17&  8.55&  2.26&     \\
163917&K0~III   &  3.32&  4928&  2.63&  1.46& +0.13&  21.4&  0.04&  0.16& $-$0.19& $-$0.28&  2.09&  3.04&  8.52&  2.56&     \\
165760&G8~III-IV&  4.64&  4962&  2.52&  1.41& $-$0.01&  13.7&  0.06&  0.25& +0.08& $-$0.27&  1.98&  2.82&  8.63&  2.65&     \\
167042&K1~III   &  5.97&  4943&  3.28&  1.07& +0.00&  20.0&  0.03&  0.01& +2.47& $-$0.28&  1.02&  1.50&  9.45&  3.32& PHS \\
167768&G3~III   &  5.99&  4895&  2.13&  1.44& $-$0.70&   9.9&  0.08&  0.39& +0.58& $-$0.30&  1.79&  2.07&  8.90&  2.68&     \\
168656&G8~III   &  4.85&  5045&  2.66&  1.30& $-$0.06&  12.1&  0.07&  0.29& $-$0.02& $-$0.25&  2.00&  2.86&  8.60&  2.66&     \\
168723&K0~III-IV&  3.23&  4972&  3.12&  1.17& $-$0.18&  52.8&  0.01&  0.06& +1.78& $-$0.27&  1.30&  1.84&  9.14&  3.15&     \\
170474&K0~III   &  5.38&  4978&  2.83&  1.29& +0.02&  13.8&  0.06&  0.30& +0.78& $-$0.27&  1.70&  2.47&  8.78&  2.88&     \\
171391&G8~III   &  5.12&  5057&  2.79&  1.23& $-$0.02&  11.2&  0.07&  0.37& +0.01& $-$0.24&  1.99&  2.84&  8.62&  2.67&     \\
174980&K0~II-III&  5.25&  5008&  2.71&  1.41& +0.10&   9.7&  0.05&  0.02& +0.17& $-$0.26&  1.94&  2.81&  8.62&  2.70&     \\
176598&G8~III   &  5.62&  5018&  2.83&  1.21& +0.03&  10.4&  0.04&  0.02& +0.68& $-$0.25&  1.73&  2.52&  8.76&  2.86&     \\
176707&G8~III   &  6.32&  4777&  2.27&  1.38& $-$0.29&   7.5&  0.07&  0.03& +0.68& $-$0.35&  1.77&  2.01&  9.03&  2.64&     \\
177241&K0~III   &  3.76&  4906&  2.70&  1.36& +0.01&  23.5&  0.03&  0.13& +0.48& $-$0.29&  1.83&  2.63&  8.71&  2.75&     \\
177249&G5~IIbCN.&  5.51&  5251&  2.55&  1.65& +0.00&   6.6&  0.07&  0.04& $-$0.44& $-$0.19&  2.15&  3.12&  8.51&  2.62&     \\
180540&K0~III   &  4.88&  4951&  2.34&  1.76& $-$0.08&   6.1&  0.14&  0.46& $-$1.66& $-$0.28&  2.67&  4.34&  8.11&  2.14&     \\
180711&G9~III   &  3.07&  4885&  2.62&  1.38& $-$0.13&  32.5&  0.01&  0.01& +0.62& $-$0.30&  1.77&  2.32&  8.91&  2.74&     \\
181276&K0~III   &  3.80&  4986&  2.78&  1.32& +0.04&  26.5&  0.02&  0.01& +0.90& $-$0.27&  1.65&  2.41&  8.81&  2.92&     \\
182694&G6.5~IIIa&  5.85&  5067&  2.63&  1.37& $-$0.04&   8.1&  0.06&  0.06& +0.32& $-$0.24&  1.87&  2.67&  8.69&  2.77&     \\
182762&K0~III   &  5.14&  4872&  2.57&  1.34& $-$0.07&  13.8&  0.05&  0.10& +0.74& $-$0.31&  1.73&  2.42&  8.82&  2.80&     \\
183491&K0~III   &  5.82&  4901&  2.63&  1.40& +0.11&   6.7&  0.11&  0.20& $-$0.24& $-$0.29&  2.11&  3.07&  8.51&  2.53&     \\
184010&K0~III-IV&  5.89&  5011&  3.17&  1.16& $-$0.14&  16.9&  0.04&  0.10& +1.93& $-$0.26&  1.23&  1.82&  9.16&  3.22&     \\
185018&G0~Ib    &  5.98&  5467&  1.85&  2.31& $-$0.10&   2.9&  0.29&  0.71& $-$2.45& $-$0.14&  2.94&  4.76&  8.06&  2.08&     \\
185194&G8~IIIvar&  5.67&  4978&  2.44&  1.54& +0.03&   6.9&  0.10&  0.19& $-$0.33& $-$0.27&  2.14&  3.09&  8.52&  2.53&     \\
185351&K0~III   &  5.17&  5006&  3.16&  1.15& +0.00&  24.6&  0.02&  0.02& +2.11& $-$0.26&  1.16&  1.76&  9.23&  3.28&     \\
185467&K0~III   &  5.97&  4937&  2.70&  1.45& +0.13&   7.9&  0.12&  0.28& +0.17& $-$0.28&  1.95&  2.83&  8.61&  2.67&     \\
185758&G0~II    &  4.39&  5535&  2.39&  1.87& +0.01&   6.9&  0.10&  0.19& $-$1.61& $-$0.13&  2.60&  4.11&  8.18&  2.38&     \\
185958&G8~II    &  4.39&  4876&  2.22&  2.08& +0.02&   7.0&  0.10&  0.19& $-$1.58& $-$0.30&  2.65&  4.33&  8.11&  2.13&     \\
186675&G8~III   &  4.89&  4953&  2.46&  1.47& $-$0.08&  11.7&  0.04&  0.04& +0.19& $-$0.28&  1.93&  2.74&  8.66&  2.67&     \\
187739&K0~III   &  5.88&  4771&  2.71&  1.03& $-$0.19&  10.5&  0.09&  0.29& +0.69& $-$0.35&  1.76&  2.01&  9.08&  2.64&     \\
188310&K0~III   &  4.71&  4802&  2.72&  1.42& $-$0.18&  16.0&  0.06&  0.10& +0.63& $-$0.33&  1.78&  2.29&  8.89&  2.69& PHS \\
188650&Fp      &  5.79&  5450&  1.79&  2.17& $-$0.67&   2.1&  0.24&  0.95& $-$3.53& $-$0.16&  3.38&  4.64&  8.07&  1.63&     \\
188947&K0~IIIvar&  3.89&  4866&  2.69&  1.35& +0.07&  23.4&  0.02&  0.09& +0.65& $-$0.31&  1.76&  2.56&  8.74&  2.78&     \\
189127&G9~III   &  6.10&  4760&  2.28&  1.41& $-$0.22&   7.1&  0.07&  0.04& +0.30& $-$0.35&  1.92&  2.31&  8.92&  2.55&     \\
192787&K0~III   &  5.70&  5025&  2.86&  1.25& $-$0.07&  10.9&  0.05&  0.20& +0.68& $-$0.25&  1.73&  2.47&  8.77&  2.86&     \\
192879&G8~III   &  5.86&  4886&  2.62&  1.37& $-$0.09&   9.6&  0.09&  0.31& +0.45& $-$0.30&  1.84&  2.47&  8.83&  2.70&     \\
192944&G8~III   &  5.30&  4981&  2.48&  1.48& $-$0.06&   6.9&  0.09&  0.18& $-$0.68& $-$0.27&  2.28&  3.41&  8.40&  2.44&     \\
192947&G6/G8~III&  3.58&  5046&  2.90&  1.32& +0.03&  30.0&  0.03&  0.12& +0.85& $-$0.25&  1.66&  2.43&  8.80&  2.93&     \\
194013&G8~III-IV&  5.30&  4906&  2.63&  1.32& $-$0.07&  13.2&  0.06&  0.05& +0.86& $-$0.29&  1.67&  2.36&  8.84&  2.86&     \\
194577&G6~III   &  5.68&  5028&  2.68&  1.34& $-$0.02&   6.0&  0.12&  0.20& $-$0.63& $-$0.25&  2.25&  3.35&  8.43&  2.47&     \\
196857&K0~III   &  5.79&  4878&  2.55&  1.44& $-$0.27&   9.9&  0.11&  0.15& +0.62& $-$0.31&  1.77&  2.15&  9.01&  2.70&     \\
199665&G6~III:  &  5.51&  4985&  2.84&  1.19& $-$0.05&  13.7&  0.05&  0.04& +1.15& $-$0.27&  1.55&  2.25&  8.90&  2.99& PHS \\
200039&G5~III   &  5.99&  4965&  2.67&  1.36& $-$0.13&   7.5&  0.07&  0.02& +0.35& $-$0.27&  1.87&  2.62&  8.70&  2.72&     \\
201381&G8~III   &  4.50&  4951&  2.77&  1.30& $-$0.04&  19.9&  0.04&  0.07& +0.93& $-$0.28&  1.64&  2.35&  8.85&  2.91&     \\
203222&G7~III:  &  5.87&  5067&  2.78&  1.29& $-$0.02&   9.7&  0.09&  0.11& +0.69& $-$0.24&  1.72&  2.49&  8.77&  2.89&     \\
203387&G8~III   &  4.28&  5244&  3.07&  1.26& +0.07&  15.1&  0.05&  0.08& +0.10& $-$0.19&  1.94&  2.79&  8.63&  2.78&     \\
204381&K0~III   &  4.50&  5100&  2.84&  1.33& $-$0.06&  18.2&  0.05&  0.09& +0.71& $-$0.23&  1.71&  2.47&  8.78&  2.90&     \\
204771&K0~III   &  5.22&  4967&  2.93&  1.26& +0.09&  14.6&  0.04&  0.16& +0.88& $-$0.27&  1.66&  2.44&  8.79&  2.91&     \\
205072&G6~III:  &  5.97&  4995&  2.72&  1.34& $-$0.14&   9.2&  0.05&  0.02& +0.76& $-$0.26&  1.70&  2.41&  8.80&  2.87&     \\
205435&G8~III   &  3.98&  5114&  3.00&  1.20& $-$0.10&  26.2&  0.02&  0.09& +0.98& $-$0.23&  1.60&  2.33&  8.85&  2.99&     \\
206356&K0~III   &  5.24&  4938&  2.80&  1.28& +0.11&  13.2&  0.06&  0.16& +0.68& $-$0.28&  1.74&  2.55&  8.74&  2.83&     \\
206453&G8~III   &  4.72&  5038&  2.43&  1.48& $-$0.38&  11.2&  0.07&  0.13& $-$0.16& $-$0.25&  2.07&  2.97&  8.53&  2.61&     \\
209396&K0~III   &  5.55&  4999&  2.81&  1.30& +0.04&  12.0&  0.07&  0.12& +0.83& $-$0.26&  1.67&  2.46&  8.79&  2.90&     \\
210354&G6~III:  &  5.58&  4793&  2.36&  1.39& $-$0.22&  11.5&  0.06&  0.05& +0.84& $-$0.34&  1.70&  1.92&  9.12&  2.70&     \\
210434&K0~III-IV&  5.98&  4949&  2.93&  1.36& +0.12&  11.6&  0.07&  0.13& +1.17& $-$0.28&  1.54&  2.29&  8.87&  2.99&     \\
210702&K1~III   &  5.93&  4967&  3.19&  1.10& +0.01&  17.9&  0.04&  0.05& +2.14& $-$0.27&  1.15&  1.68&  9.28&  3.25& PHS \\
210807&G8~III   &  4.79&  5071&  2.58&  1.57& $-$0.10&   8.6&  0.05&  0.28& $-$0.81& $-$0.24&  2.32&  3.50&  8.37&  2.44&     \\
211391&G8~III-IV&  4.17&  4909&  2.57&  1.36& +0.09&  17.0&  0.04&  0.10& +0.23& $-$0.29&  1.92&  2.78&  8.64&  2.68&     \\
211434&G6~III   &  5.75&  5082&  2.70&  1.37& $-$0.26&   9.6&  0.09&  0.14& +0.51& $-$0.24&  1.79&  2.53&  8.73&  2.83&     \\
211554&G8~III   &  5.88&  5043&  2.41&  1.63& +0.05&   4.5&  0.12&  0.69& $-$1.55& $-$0.25&  2.62&  4.26&  8.12&  2.21&     \\
212271&K0~IIICN.&  5.53&  5002&  2.90&  1.21& +0.10&  12.0&  0.06&  0.16& +0.77& $-$0.26&  1.70&  2.50&  8.76&  2.89&     \\
212320&G6~V     &  5.92&  5075&  2.59&  1.46& $-$0.11&   7.1&  0.13&  0.17& +0.01& $-$0.24&  1.99&  2.84&  8.61&  2.68&     \\
212430&K0~III   &  5.76&  4954&  2.56&  1.39& $-$0.17&   6.0&  0.13&  0.17& $-$0.52& $-$0.28&  2.22&  3.17&  8.49&  2.45&     \\
212496&G9~III   &  4.42&  4710&  2.43&  1.22& $-$0.33&  19.2&  0.03&  0.17& +0.67& $-$0.38&  1.78&  1.85&  9.12&  2.57&     \\
213789&G6~III   &  5.88&  5010&  2.73&  1.37& $-$0.06&   7.3&  0.12&  0.07& +0.14& $-$0.26&  1.95&  2.77&  8.64&  2.69&     \\
213930&G8~III-IV&  5.72&  5011&  2.87&  1.34& +0.12&   9.6&  0.06&  0.34& +0.29& $-$0.26&  1.89&  2.75&  8.65&  2.75&     \\
213986&K1~III   &  5.97&  4928&  2.83&  1.27& +0.08&   9.7&  0.09&  0.14& +0.75& $-$0.29&  1.72&  2.50&  8.76&  2.85&     \\
214567&G8~II    &  5.84&  4989&  2.69&  1.33& $-$0.21&   8.6&  0.09&  0.09& +0.41& $-$0.27&  1.84&  2.57&  8.71&  2.75&     \\
214878&B8~V     &  5.94&  5041&  2.85&  1.29& +0.04&   9.5&  0.06&  0.34& +0.49& $-$0.25&  1.80&  2.62&  8.71&  2.82&     \\
215030&G9~III   &  5.93&  4731&  2.41&  1.25& $-$0.49&  10.1&  0.06&  0.06& +0.89& $-$0.37&  1.69&  1.83&  9.24&  2.67&     \\
215373&K0~III   &  5.11&  5007&  2.69&  1.39& +0.10&  11.9&  0.05&  0.05& +0.44& $-$0.26&  1.83&  2.66&  8.69&  2.79&     \\
215721&G8~III   &  5.24&  4829&  2.23&  1.39& $-$0.48&  12.3&  0.07&  0.10& +0.58& $-$0.33&  1.80&  1.95&  9.07&  2.62&     \\
215943&G8~III:  &  5.82&  4878&  2.68&  1.33& $-$0.04&   9.0&  0.08&  0.06& +0.52& $-$0.30&  1.81&  2.45&  8.84&  2.72&     \\
216131&M2~III   &  3.51&  5000&  2.69&  1.24& $-$0.05&  27.9&  0.03&  0.06& +0.68& $-$0.26&  1.73&  2.49&  8.77&  2.85&     \\
217264&K1~III:  &  5.43&  4946&  2.80&  1.27& +0.12&  11.6&  0.08&  0.07& +0.69& $-$0.28&  1.74&  2.55&  8.74&  2.84&     \\
217703&K0~III   &  5.97&  4890&  2.91&  1.16& $-$0.17&  13.0&  0.06&  0.10& +1.44& $-$0.30&  1.44&  1.98&  9.05&  3.00&     \\
218527&G8~IV    &  5.42&  4935&  2.57&  1.33& $-$0.34&  11.6&  0.11&  0.07& +0.68& $-$0.29&  1.74&  2.11&  9.03&  2.75&     \\
219139&G5~III:  &  5.85&  4860&  2.50&  1.38& $-$0.19&   9.7&  0.08&  0.07& +0.72& $-$0.31&  1.74&  2.29&  8.88&  2.76&     \\
219615&G7~III   &  3.70&  4802&  2.25&  1.37& $-$0.62&  24.9&  0.04&  0.06& +0.62& $-$0.34&  1.79&  1.67&  9.20&  2.55&     \\
219945&K0~III   &  5.44&  4874&  2.61&  1.36& $-$0.10&   9.9&  0.06&  0.18& +0.25& $-$0.31&  1.92&  2.57&  8.77&  2.63&     \\
221345&K0~III   &  5.22&  4813&  2.63&  1.43& $-$0.24&  13.1&  0.05&  0.13& +0.67& $-$0.33&  1.76&  2.20&  8.93&  2.70& \\
222093&K0~III   &  5.66&  4853&  2.56&  1.38& $-$0.12&  11.5&  0.08&  0.10& +0.86& $-$0.31&  1.68&  2.28&  8.89&  2.81&     \\
222387&G8~III   &  5.98&  5055&  2.81&  1.22& $-$0.11&   7.8&  0.08&  0.31& +0.12& $-$0.24&  1.95&  2.79&  8.63&  2.70&     \\
222574&G2~Ib/II &  4.82&  5523&  1.99&  2.20& +0.04&   5.1&  0.16&  0.10& $-$1.75& $-$0.13&  2.65&  4.23&  8.13&  2.34&     \\
223252&G8~III   &  5.49&  5031&  2.72&  1.34& $-$0.03&  11.2&  0.08&  0.10& +0.63& $-$0.25&  1.75&  2.52&  8.76&  2.85&     \\
224533&G9~III   &  4.88&  5030&  2.73&  1.28& $-$0.01&  14.6&  0.06&  0.10& +0.60& $-$0.25&  1.76&  2.54&  8.75&  2.84&     \\
\end{longtable}


\newpage

\begin{thebibliography}{}
%
\bibitem[]{}
  Alonso, A., Arribas, S., \& Mart\'{\i}nez-Roger, C. 1999,
  A\&AS, 140, 261
\bibitem[]{}
  Arenou, F., Grenon, M., \& G\'{o}mez, A. 1992, A\&A, 258, 104
\bibitem[]{}
  Boss, A. P. 1997, Science, 276, 1836
\bibitem[]{}
  da Silva, L., et al. 2006, A\&A, 458, 609
\bibitem[]{}
  de Medeiros, J. R., \& Mayor, M. 1999, A\&AS, 139, 433
\bibitem[]{}
  ESA 1997, The Hipparcos and Tycho Catalogues, ESA SP-1200
\bibitem[]{}
  Gonzalez, G. 2003, Reviews of Modern Physics, 75, 101
\bibitem[]{}
  Gray, D. F. 1988, Lectures on Spectral-Line Analysis: F, G, and K stars
  (Arva, Ontario: The Publisher)
\bibitem[]{}
  Gray, D. F. 1989, ApJ, 347, 1021
\bibitem[]{}
  Gray, D. F. 2005, The Observation and Analysis of Stellar Photospheres,
  3rd ed. (Cambridge: Cambridge University Press)
\bibitem[]{}
  Hatzes, A. P., et al. 2006, A\&A, 457, 335
\bibitem[]{}
  Hayashi, C., Nakazawa, K., \& Nakagawa, Y. 1985,
  in Protostars and Planets II (A86-12626 03-90) 
  (Tucson, Arizona: University of Arizona Press), p.1100 
\bibitem[]{}
  Hekker, S., \& Mel\'{e}ndez, J. 2007, A\&A, 475, 1003
\bibitem[]{}
  Ibukiyama, A., \& Arimoto, N. 2002, A\&A, 394, 927
\bibitem[]{}
  Ida, S., \& Lin, D. N. C 2004, ApJ, 616, 567
\bibitem[]{}
  Johnson, J. A., et al. 2007, ApJ, 665, 785
\bibitem[]{}
  Kraft, R. P. 1994, PASP, 106, 553
\bibitem[]{}
  Kurucz, R. L. 1992, in The Stellar Populations of Galaxies,
  Proc. IAU Symp. 149, eds. B. Barbuy \& A. Renzini (Dordrecht: Kluwer), p.225
\bibitem[]{}
  Kurucz, R. L. 1993, Kurucz CD-ROM  No.13, Atlas 9 Stellar 
  Atmosphere Programs and 2 km~s$^{-1}$ Grid 
  (Cambridge: Smithsonian Astrophysical Observatory) 
  [available at $\langle$http://kurucz.harvard.edu/cdroms.html$\rangle$]
\bibitem[]{}
  Kurucz, R. L., \& Bell, B. 1995, Kurucz CD-ROM No.23,
  Atomic Line List (Cambridge: Smithsonian Astrophysical Observatory) 
  [available at $\langle$http://kurucz.harvard.edu/cdroms.html$\rangle$]
\bibitem[]{}
  Kurucz, R. L., Furenlid, I., Brault, J., \& Testerman, L. 1984,  
  Solar Flux Atlas from 296 to 1300 nm
  (Sunspot, New Mexico: National Solar Observatory) 
  [available at $\langle$http://kurucz.harvard.edu/sun.html$\rangle$]
\bibitem[]{}
  Lejeune, T., \& Schaerer, D. 2001, A\&A, 366, 538
\bibitem[]{}
  Liu, Y.-J., et al. 2008, ApJ, 672, 553
\bibitem[]{}
  Luck, R. E., \& Heiter, U. 2007, AJ, 133, 2464
\bibitem[]{}
  Luck, R. E., \& Lambert, D. L. 1985, ApJ, 298, 782
\bibitem[]{}
  Massarotti, A., Latham, D. W., Stefanik, R. P., \& Fogel, J. 2008,
  AJ, 135, 209
\bibitem[]{}
  McWilliam, A. 1990, ApJS, 74, 1075
\bibitem[]{}
  Pasquini, L., D\"{o}llinger, M. P., Weiss, A., Girardi, L., 
  Chavero, C., Hatzes, A. P., da Silva, L., \& Setiawan, J. 2007,
  A\&A, 473, 979 
\bibitem[]{}
  Reffert, S., Quirrenbach, A., Mitchell, D. S., Simon, A., Hekker, S.,
  Fischer, D. A., Marcy, G. W., \& Butler, R. P. 2006, AJ, 652, 661
\bibitem[]{}
  Sato, B., et al. 2003, ApJ, 597, L157
\bibitem[]{}
  Sato, B., et al. 2007, ApJ, 661, 527
\bibitem[]{}
  Sato, B., et al. 2008, PASJ, in press
\bibitem[]{}
  Takeda, Y. 1995, PASJ, 47, 287
\bibitem[]{}
  Takeda, Y. 2007, PASJ, 59, 335
\bibitem[]{}
  Takeda, Y., et al. 2005a, PASJ, 57, 13
\bibitem[]{}
  Takeda, Y., \& Honda, S. 2005, PASJ, 57, 65
\bibitem[]{}
  Takeda, Y.,  Kawanomoto, S., Honda, S., Ando, H., \& Sakurai, T. 2007, 
  A\&A, 468, 663
\bibitem[]{}
  Takeda, Y.,  Kawanomoto, S., \& Sadakane, K. 1998, 
  PASJ, 50, 97
\bibitem[]{}
  Takeda, Y., Ohkubo, M., \& Sadakane, K.  2002, 
  PASJ, 54, 451
\bibitem[]{}
  Takeda, Y., Ohkubo, M., Sato, B., Kambe, E., \& Sadakane, K.  2005b, 
  PASJ, 57, 27 [Erratum: 57, 415]
\bibitem[]{}
  Takeda, Y., Sato, B., Kambe, E., Izumiura, H., Masuda, S., 
  \& Ando, H. 2005c, PASJ, 57, 109 (Paper~I)
\bibitem[]{}
  Takeda, Y., \& Takada-Hidai, M. 1994, PASJ, 46, 395
\bibitem[]{}
  Udry, S., \& Santos, N. 2007, ARA\&A, 45, 397
\bibitem[]{}
  Zhao, G., Qiu, H. M., \& Mao, S. 2001, ApJ, 551, L85
\end{thebibliography}
\end{document}